\documentclass[12pt]{article}
\usepackage{setspace}
\usepackage{multirow}
\usepackage{subfigure}
\usepackage{amsmath,amssymb,amsfonts,mathrsfs}
\usepackage{natbib}
\usepackage{graphicx, epstopdf}
\usepackage{fullpage}

\usepackage{graphicx}
\usepackage{graphics}
\usepackage{multimedia}
\usepackage[english]{babel}
\usepackage[latin1]{inputenc}
\usepackage{times}
\usepackage[T1]{fontenc}
\usepackage{pgf,pgfarrows,pgfnodes,pgfautomata,pgfheaps,pgfshade}
\usepackage[latin1]{inputenc}
\usepackage{colortbl}
\usepackage{color}
\usepackage{xcolor}
\usepackage[english]{babel}
\usepackage{bm}
\usepackage{bbm}
\usepackage{multimedia}
\usepackage{subfig}
\usepackage{multicol}
\usepackage{pdfpages}
\usepackage{enumerate}
 \usepackage{epstopdf}

\newcommand{\Cov}{\mathop{\rm {\mathbb C}ov}\nolimits}%
\newcommand{\Cor}{\mathop{\rm {\mathbb C}or}\nolimits}%
\newcommand{\cov}{\mathop{\rm {\mathbb C}ov}\nolimits}%
\newcommand{\ex}{{\mathbb E}}

\DeclareMathOperator*{\argmin}{ arg\,min}

\xdefinecolor{DarkBlue}{rgb}{0,0,0.65}

\usepackage{bm}
\usepackage{xr-hyper}
\usepackage{color}
\usepackage{graphicx}
\newcommand{\E}{{\mathbb E}}

\usepackage{xcolor}

\usepackage{algorithm}
\usepackage{algpseudocode}
\usepackage{pifont}
\usepackage[titletoc,title]{appendix}
\title{Exploratory Analysis of High Dimensional Time Series with Applications to Multichannel  Electroencephalograms}
\date{}
\author{Yuxiao Wang$^1$, Chee-Ming Ting$^2$ and Hernando Ombao$^{1,3}$}
\footnotetext[1]{Department of Statistics, University of California, Irvine, U.S.A.}
\footnotetext[2]{Department of Biomedical Engineering, Universiti Teknologi, Malaysia}
\footnotetext[3]{Corresponding author: hombao@uci.edu}

\begin{document}
\maketitle
\abstract{
In this paper, we address the the major hurdle of high dimensionality in EEG analysis by extracting the optimal lower dimensional representations. Using our approach, connectivity between regions in a high-dimensional brain network is characterized through the connectivity between region-specific factors. The proposed approach is motivated by our observation that electroencephalograms (EEGs) from channels within each region exhibit a high degree of multicollinearity and synchrony. These observations suggest that it would be sensible to extract summary factors for each region. We consider the general approach for deriving summary factors which are solutions to the criterion of squared error reconstruction. In this work, we focus on two special cases of linear auto encoder and decoder. In the first approach, the factors are characterized as instantaneous linear mixing of the observed high dimensional time series. In the second approach, the factors signals are linear filtered versions of the original signal which is more general than an instantaneous mixing. This exploratory analysis is the starting point to the multi-scale factor analysis model where the concatenated factors from all regions are represented by vector auto-regressive model that captures the connectivity in high dimensional signals.  We performed evaluations on the two approaches via simulations under different conditions.  The simulation results provide insights on the performance and application scope of the methods. We also performed exploratory analysis of EEG recorded over several epochs during resting state.  Finally, we implemented these exploratory methods in a Matlab toolbox XHiDiTS available from {https://goo.gl/uXc8ei}.
}

\newpage

\section{Introduction}
\begin{figure}[h]
    \centering
    \includegraphics[width=1.0\textwidth]{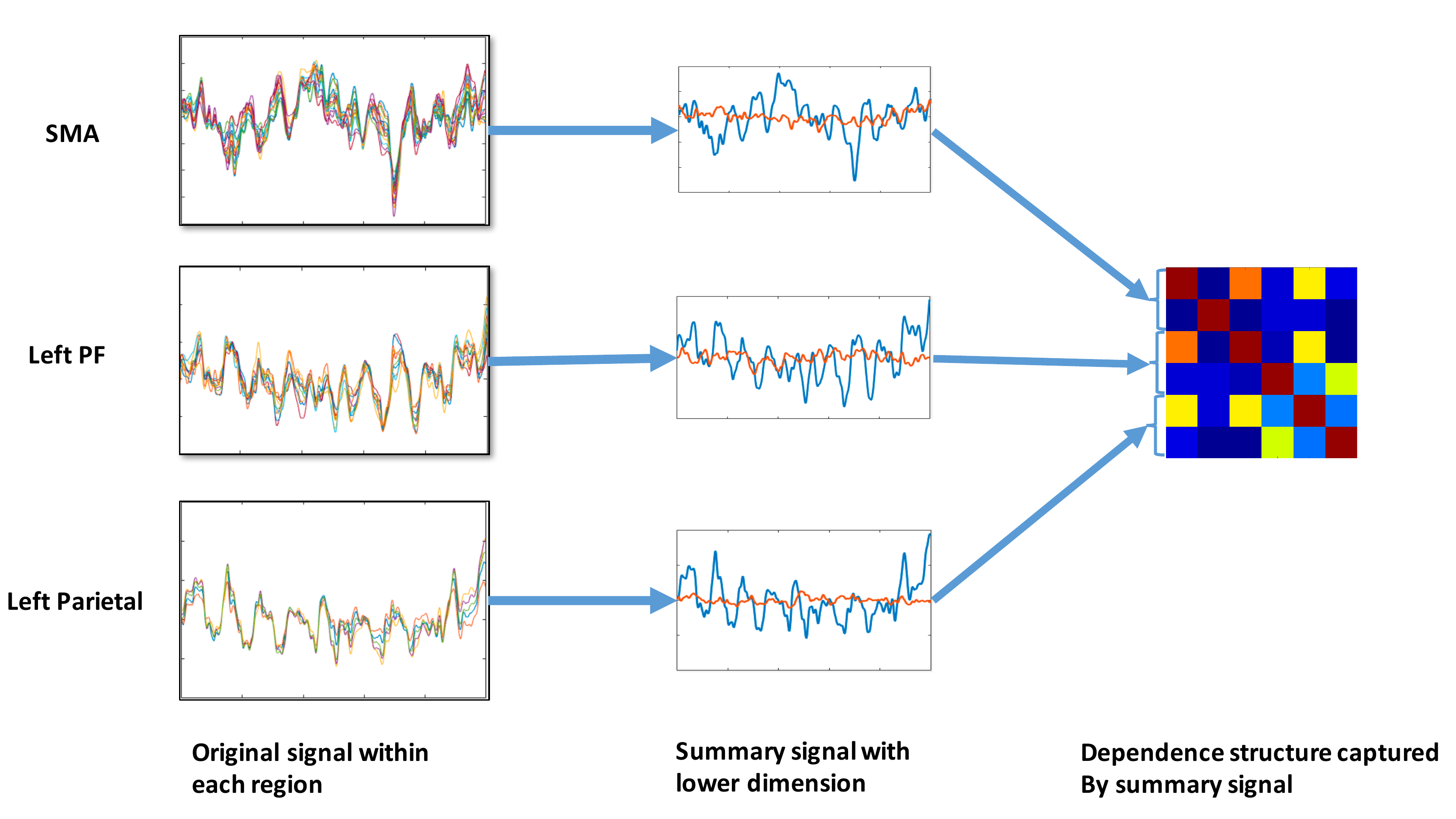}
    \caption{Illustration of the modeling procedure. The goal is to characterize the dependence between three different regions: SMA, left Pre-frontal cortex and left parietal.  As the first step, summary factors are obtained. Then the dependence between the summary factors are computed. }
    \label{fig:demo_factor_model}
\end{figure}

\begin{figure}[!htb]%
    \centering
    \subfigure[Plot of EEGs on a 2-d scalp]{\includegraphics[width=0.8\textwidth, height=8cm]{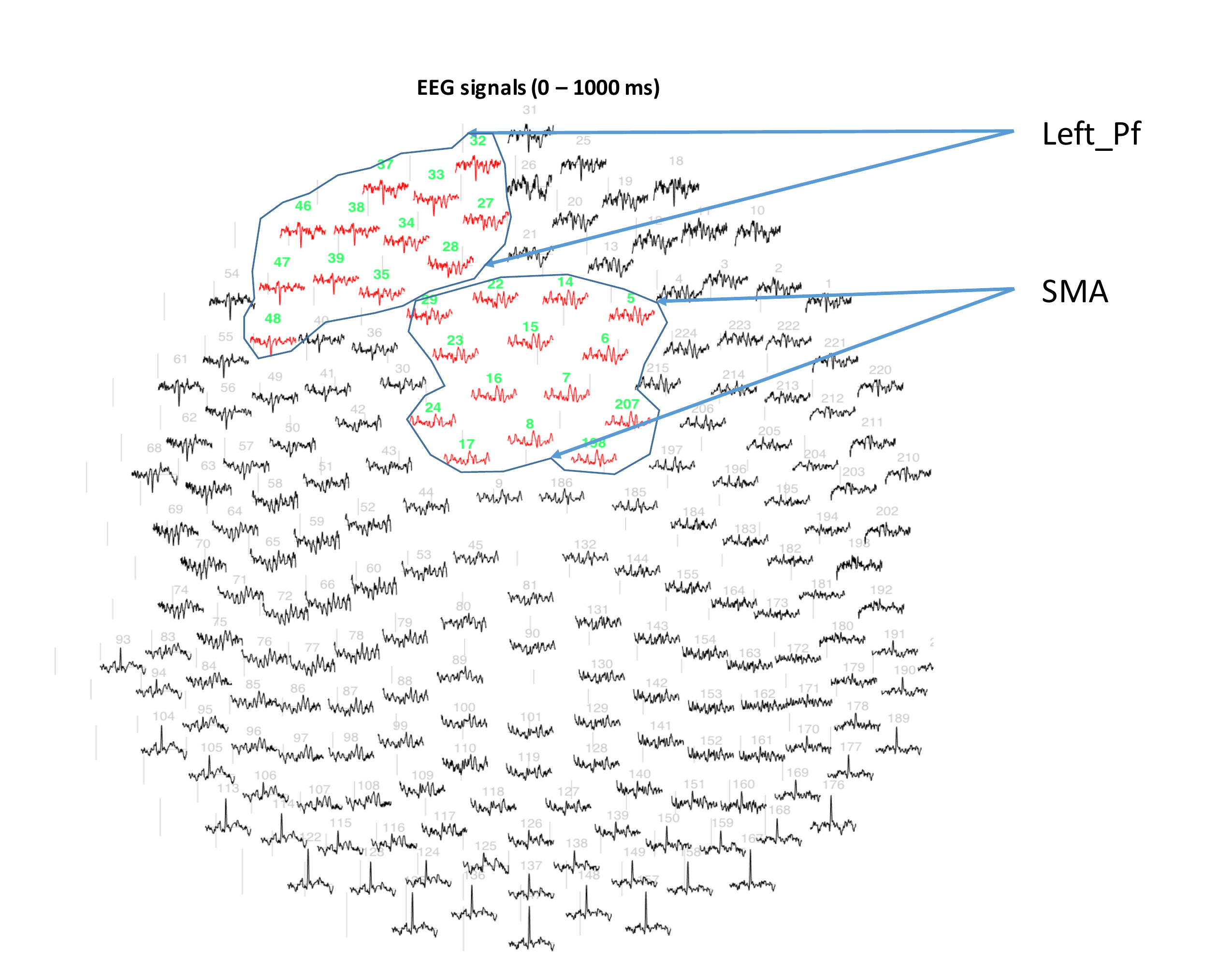} }\\
    \subfigure[Plot of EEGs in SMA region]{{\includegraphics[width=0.48\textwidth, height=5cm]{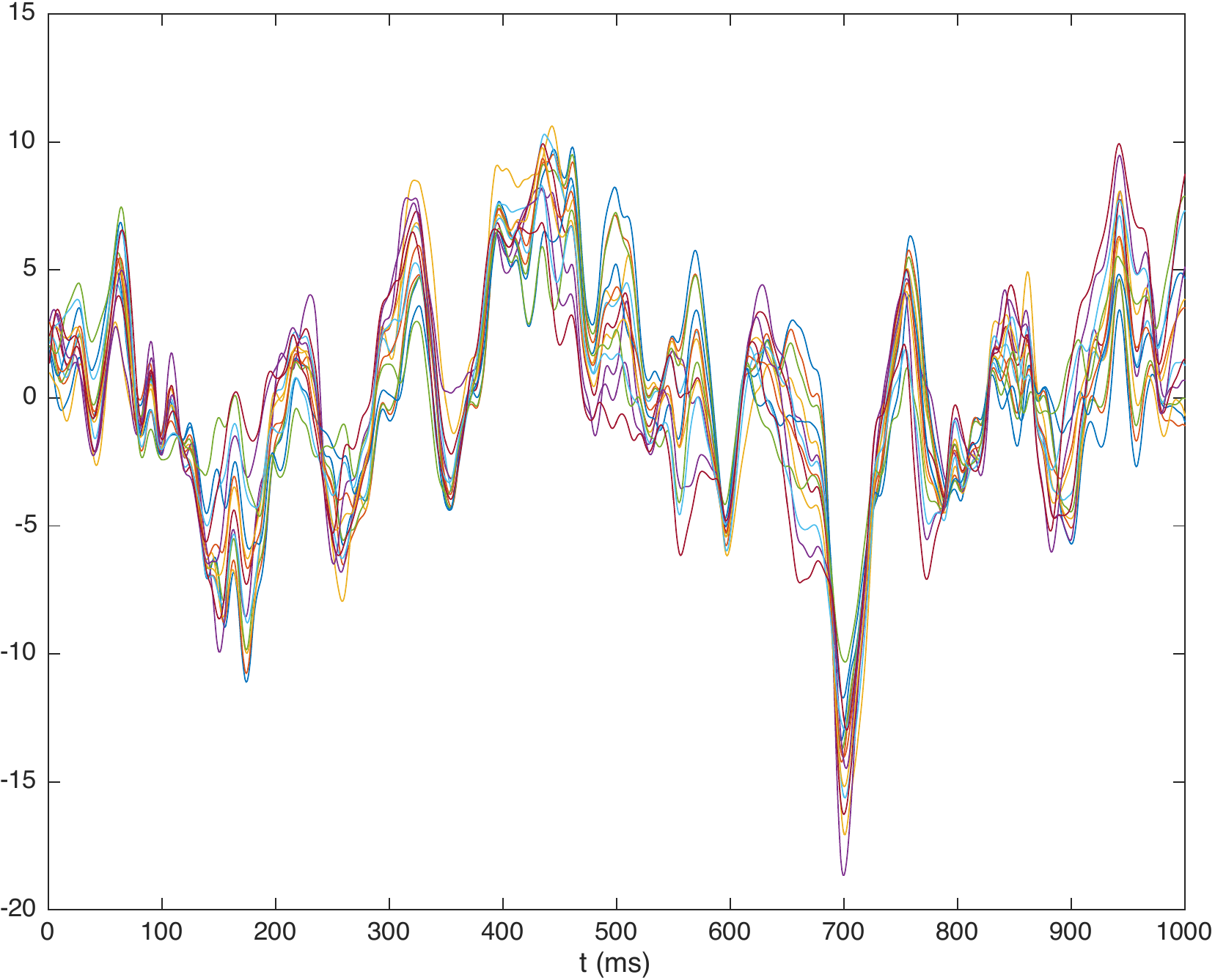} }}
    \subfigure[Plot of EEGs in left Pf region]{{\includegraphics[width=0.48\textwidth, height=5cm]{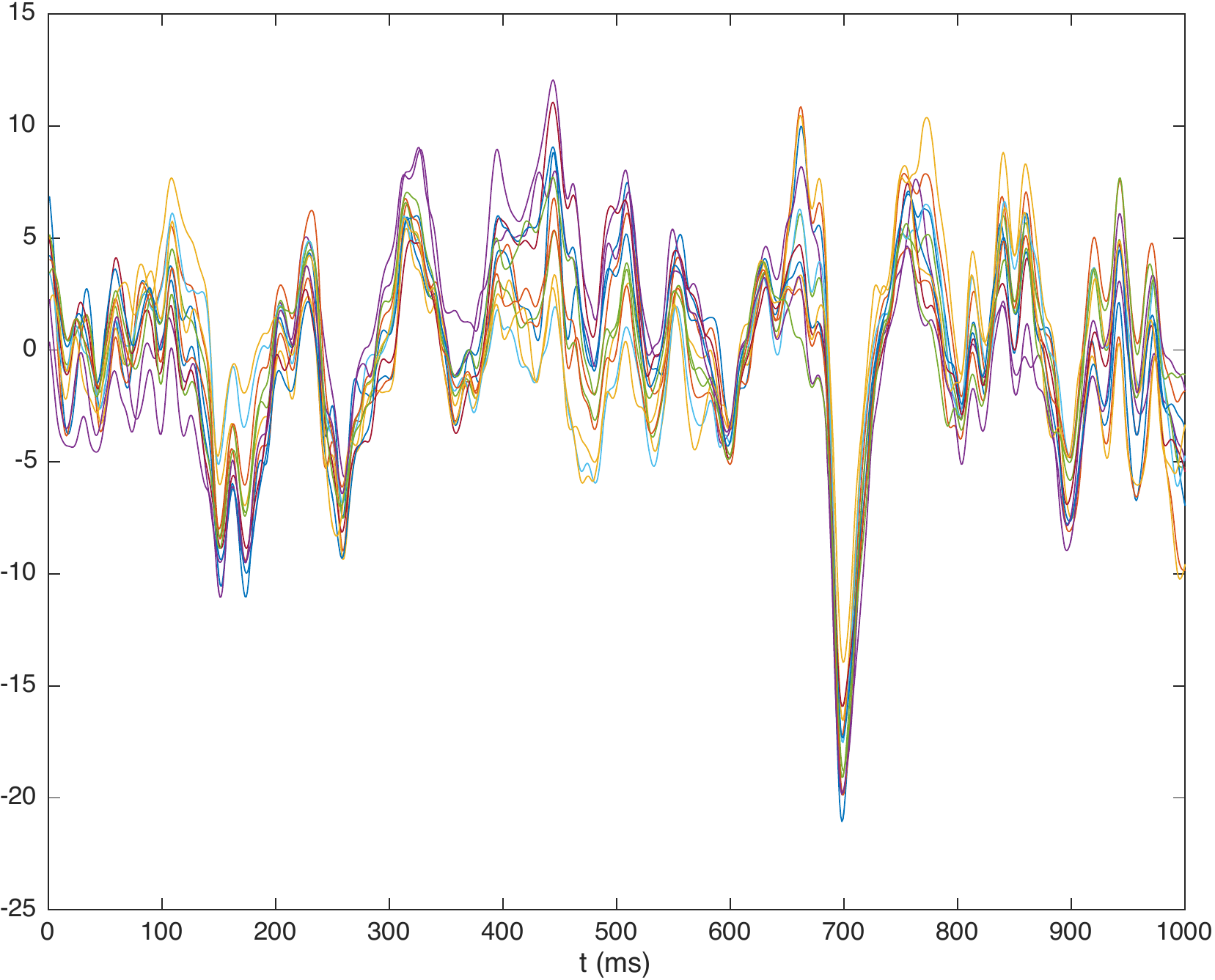} }}
    \caption{Visualization of EEGs at SMA region (bottom left) and left Pre-frontal region (bottom right) for one trial (1000 ms). These two plots combined with the visualization on the 2-d scalp (top) show that the EEGs within a region are highly correlated.}%
    \label{fig:eeg_two_regions}%
\end{figure}

Neuronal populations behave in a coordinated manner in order to execute learning, memory retention and even during resting state. In fact, disruptions in connectivity between brain regions is associated with a number of neurological diseases such as schizophrenia,  obsessive compulsive disorder and Alzheimer's disease. In this project, we shall investigate connectivity between brain regions using electroencephalograms (EEGs) which indirectly measure cortical neuronal activity. The key challenge in estimating connectivity brain networks is the high dimensionality of these signals. Another is the high multicollinearity between these channels that is primarily due to the spatial filtering and volume conduction. Finally,  since these EEG signals are recorded over several epochs in an experiment, it is vital to understand how connectivity patterns may change across the experiment. Empirical inspection of the EEGs in each region show a high degree of multicollinearity (see Figure \ref{fig:eeg_two_regions}). Therefore it is sensible to perform dimensionality reduction at region level. One approach to modeling the connectivity is to first derive low dimensional representation or simpler structure of the original signal and secondly, analyze the connectivity structure or other properties in the lower dimensional space. The approach is illustrated in Figure \ref{fig:demo_factor_model}. In this paper, we focus on the first step of the approach and study two models that can be used to reduce the dimensionality of the high dimensional EEG time series. This is a first step towards dimension reduction which could lead to better statistical modeling and inference. It could also help to identify any potential irregularities in the signals (e.g. outliers, non-stationarities across epochs).  

The remainder of the paper is organized as follows. Section \ref{sec:dimension_reduction} describes the methods for dimensionality reduction for time series data. Section \ref{sec:simulation} presents the evaluation results of the models based on simulated data. Section \ref{sec:data_analysis} presents the results of an exploratory analysis on a real resting-state EEG dataset followed by a conclusion in Section \ref{sec:conclusion}.

\section{Dimensionality reduction of time series}\label{sec:dimension_reduction}
In this section, we describe the approach to performing exploratory analysis of high-dimensional EEG time series. Let $R$ be the number of regions on the scalp area. The EEG signals at region $r \in \{1, \dots, R\}$ can be represented using ${\bf z}_r(t)$, where the dimension of ${\bf z}_r(t)$ is $n_r$, which is equal to the number of channels within region $r$. EEG signals within a region appear to be highly correlated, which indicates that the variability can be well captured by a more compressed time series (the factors) of lower dimension. Besides that, modeling through low-dimensional representation or simplified structure has advantages in the sense that (1.) it takes advantage of the structure of data, e.g. high correlation for EEG signals within the same region, and hence the low dimensional embedding approach serves similar purpose as imposing regularization; (2.) it enables the modeling training using a relatively small training sample size. Problem with reduced dimension enables us to train a less constrained parametric model, for example, allowing larger lags in VAR model allows us to capture more complex dependence structure;  and (3.) models with reduced size can be trained much faster, especially when the training time is higher than quadratic order of the dimension. For example, in computing partial coherence,  inversion operation is far less painful for matrix of lower dimension.

A substantial amount of research has been done on dimensionality reduction of high dimensional time series. Such methods include frequency domain approaches \citep{brillinger1964frequency, brillinger2001time}; dynamic factor models and state space approach with low dimensional state \citep{durbin2001time, harvey1990forecasting, lam2012factor}; canonical analysis \citep{box1977canonical} and principal component analysis \citep{wang2016modeling}. Here, we build on these foundations in two directions: we develop exploratory methods that handle multiple-epochs (rather than just a single epoch) and then package these into a toolbox that we hope would help neuroscientists to use a more data-adaptive approach to investigating connectivity in high dimensional brain signals.

In this paper, we will be mainly focusing on two classes of approaches based on principal component analysis. In this section, we derive factor activities for each region $r$, denoted by ${\bf f}_r(t)$, which has a lower dimension $m_r$ comparing to the original space, i.e. $m_r$ $\ll$ $n_r$.  For consistency, we use lower case letter for scalar number, bold lowercase letter for column vector, and uppercase letter for matrix. For convenience purpose, we dropped the subscript $r$ when dealing with the time series signals.
\subsection{Auto encoder for time series data}
The auto encoder algorithm is a general approach to learning representations of the input data (in this case the high dimensional time series). The algorithm was first introduced by \citep{rumelhart1985learning}. It can be used to reduce the dimension of the time series via learning a low dimensional representation. The algorithm consists of two parts, the encoder and the decoder. The encoder function, defined as $F_{en}: Z \rightarrow {\bf f}$,  is a transformation from the original high dimensional time series $Z$ into some low dimensional space. The decoder function, defined as $F_{de}: {\bf f} \rightarrow Z$ is a transformation from the encoded (low-dimensional space) to the original high dimensional space. In this paper, we will consider a special case of encoders and decoders which as linear transformations (instantaneous mixing or filtered versions) of the original time series. Denote the parameters of the encoder and the decoder to be $\Theta$, then the best encoder and the decoder are the ones that minimize the reconstruction error, i.e.,
\begin{equation}
\widehat{\Theta} = \argmin_{\Theta} ||Z - F_{de}(F_{en}(Z))||^2_F
\end{equation}
where the $|| E ||_F$ is the Frobenius norm which is defined as $|| E ||_F = \sqrt{\text{Trace}(EE^T)}$.
For the time series data ${\bf z}(t)$ generated from distribution $P(Z)$ where $Z = [{\bf z}(1)', \dots, {\bf z}(T)']'$, we consider the expected loss, that is, the best parameters are computed as
\begin{equation}
\widehat{\Theta} = \argmin_{\Theta} \E_{P(Z)}||Z - F_{de}(F_{en}(Z))||^2_F
\end{equation}
For linear encoder and decoders, i.e., both of $F_{en}(Z)$ and $F_{de}({\bf f})$ are linear functions,  the solution is strongly related to principal component analysis. In this paper, we will focus on two types of linear encoders and decoders: the first is the instantaneous mixing encoder and the second is a linear filter of the time series and thus captures the entire temporal dynamics of the time series. 

\subsection{Method 1: The factor is an instantaneous linear mixture of the time series ${\bf z} (t)$}
For a given time series process ${\bf z}(t) \in \mathbb{R}^{n}$, we will consider the problem of learning a representation of lower dimension. The optimal representation is considered as the one that gives the best reconstruction accuracy. In this approach, we derive the factor ${\bf f}(t) \in \mathbb{R}^{m}$ using the instantaneous linear mixture of the original time series ${\bf z} (t)$, as described in Equation \ref{eq:approach_1}. The dimensionality reduction can be achieved when $m$ is smaller than $n$.
\begin{equation}\label{eq:approach_1}
{\bf f}(t) = {\bf A}^T {\bf z}(t) 
\end{equation}
For the purposes of keeping the parameters identifiable, we shall assume that (a.)  ${A}^T {A} = I_{m}$ and (b.) $\Cov[{\bf f}(t)]$ is a diagonal matrix, i.e., the factors are uncorrelated.
We reconstruct ${\bf z}(t)$ using the instantaneous linear mixture of ${\bf f}(t)$ in the form of $\widehat{\bf z}(t) = {B} {\bf f}(t)$. The goal is to find ${A}$ and ${B}$ that minimize the reconstruction error defined in Equation \eqref{eq:approach_1_error_function}.
\begin{equation}\label{eq:approach_1_error_function}
{L}( A, B ) = \text{Trace} ( \ex \left [ {\bf z}(t) \  - \ {\widehat{\bf z} (t)} \right ] \ \left [ {\bf z}(t)  \ - \ {\widehat{\bf z}(t)} \right ]^T )
\end{equation}
The solution can be derived using the following two steps.
\begin{itemize}
\item Step 1. Compute the eigenvalues-eigenvectors of $\Sigma^{\bf z}(0)$ as $\{ (\lambda_{s}, {\bf e}_{s} )\}_{s=1}^{n}$ where $\lambda_{1} > \ldots, > \lambda_{n}$ and $\| {\bf e}_{s} \| = 1$. When $\Sigma^{\bf z}(0)$ is not known, we use an estimator instead, which can be computed as $\widehat{\Sigma}^{\bf z}(0) = \frac{1}{T}\sum_{t=1}^T {\bf z}(t) {\bf z}(t)^T$ assuming ${\bf z}(t)$ has zero mean.
\item Step 2. The solution can be represented by \[
\widehat{{A}}  = \widehat{{B}} = [{\bf e}_{1}, \ldots, {\bf e}_{m}]
\ \ \ {\mbox{and}} \ \ \  
\widehat{\bf f}(t) = \widehat{{A}}^T {\bf z}(t) \ .
\]
\end{itemize}
The solution is closely related to principal components analysis (PCA) of the covariance matrix of the input signals at the zero lag, i.e., $\Sigma^{\bf z}(0) = \cov ({\bf z}(t), {\bf z}(t))$. It is the one that accounts for the most of the variation of the time series, among all the instantaneous linear projections with the same dimension.

\subsection{Method 2: The factor is a linear filter of the time series ${\bf z}(t)$}
Alternative to approach 1, if we restrict the form of the representation to linear functions of all ${\bf z}(t) \in \mathbb{R}^n$ rather than merely an instantaneous linear mixture,  the lower dimensional representation denoted by ${\bf f}(t) \in \mathbb{R}^m$ can be written as 
\begin{equation}\label{eq:compression}
{\bf f}(t) = \sum_{h=\infty}^{\infty} A(h)^T {\bf z}(t-h)
\end{equation}
where $A(h) \in \mathbb{C}^{n \times m}$ with $m < n$, and ${f}_i(t)$ and ${f}_j(t)$ has zero coherency for $i \neq j$.
We consider the reconstruction of ${\bf z}(t)$ using linear function of ${\bf f}(t)$ in the following form
\begin{equation}\label{eq:reconstruction}
\widehat{\bf z}(t) = \sum_{j=-\infty}^{\infty} B(j) {\bf f}(t-j)
\end{equation} where $B(j) \in \mathbb{C}^{n \times m}$ is the transformation coefficient matrix.
The reconstruction error (loss function) is defined by the expected squared loss. That is
\begin{eqnarray}
L(\{A(h)\}, \{B(j)\}) &=& \text{Trace} ( \ex \left [ {\bf z}(t) \  - \ {\widehat{\bf z} (t)} \right ] \ \left [ {\bf z}(t)  \ - \ {\widehat{\bf z}(t)} \right ]^T )\\
&=& \text{Trace}(\Cov[\widehat{\bf z}(t) - {\bf z}(t)])
\end{eqnarray}
The best transformation is defined as the $A(h)$ and $B(j)$ values that minimize $L(\{A(h)\}, \{B(j)\})$
\begin{equation} \label{eq:loss_function_2}
\{\widehat{A}(h)\}, \{\widehat{B}(j)\} = \argmin_{\{A(h)\}, \{B(j)\}} L(\{A(h)\}, \{B(j)\})
\end{equation}
The solution to the criterion in Equation \ref{eq:loss_function_2} is obtained via principal components analysis of the spectral matrix which is described in detail in Algorithms Algorithms (\ref{algo:spectrum}) and (\ref{algo:pca_freq}). Note that this dimension reduction procedure was originally described in \citep{brillinger1964frequency}.

\subsection{Comparison of the two linear encoders}
Both encoder methods are based on projecting the original high dimensional signal onto a space of lower dimension. Both methods are similar in the sense that the factors are constrained to be linear functions of the original signal. However they differ in this respect: method 1 produces factor ${\bf f}(t)$ which only explicitly depends on the signal at time $t$. Under this approach, temporal dynamics of ${\bf z}(t)$ is ignored. The second method gives factors which are low dimensional filtered versions of the original signal. The factors at time point $t$ is obtained by using {\it all} data points ${\bf z}(t \pm \ell)$. Thus it captures temporal dynamics and lead-lag relationships in the original time series. We note here that the first method is a special case of the second. In fact, by constraining $A(h) = 0$ and $B(j) = 0$ for all $h \ne 0$ and $j \ne 0$ then the linear filtered series is reduced to the instantaneously-mixed signal. 

The key advantage of the second method is that it is likely to give lower reconstruction error because it uses all the information about the signal. The first method is particularly problematic when there is some lead-lag relationships between the original signals which could be completely washed out with the simplistic instantaneous mixing. It is also supported by the simulation results, where the second model has better performance when the time series has time shift in some channels (Figure \ref{fig:simu_4} and \ref{fig:simu_5}). 

In terms of computational complexity, model 2 needs to compute the eigenvalue decomposition of all the frequency matrices while model 1 only needs to decompose the zero-lag covariance matrix. That means the second model requires more computational resources (in both space and time), comparing to model 1. It would be helpful to identify the model with suitable complexity for the problem.  In the simulation study, we applied the two models on time series data generated from different distributions to gain better understanding of their performance.

\section{Simulation}\label{sec:simulation}
In this section, we apply the algorithms on simulated time series of various properties and evaluate the performance. The goal of the simulation is to provide comprehensive evaluation of the performance of the models including application scope, capability and computational complexity. In particular, we perform simulations to analyze the performance of the approaches in terms of reconstructing the original time series. The step of the simulations are described as follows.
\begin{itemize}
\item Step 1: model training.  Generate training time series data $Z_{tr}$ from distribution $F(Z)$ and fit the model using $Z_{tr}$. 
\item Step 2. model evaluation. Generate $K$ iid test datesets $Z_{te_1}, \dots, Z_{te_K}$. For each test dataset $Z_{te}$, we evaluate the model using the normalized reconstruction error in the form of $\frac{||Z_{te} - \widehat{Z}_{te}||_F^2}{||Z_{te}||_F^2}$, where the reconstructed time series $\widehat{Z}_{te}$ is computed by applying the trained model obtained in Step 1. The mean and the standard deviation of the test error are computed and compared across models.
\end{itemize}
We performed multiple simulations using data generated from different distributions $F(Z)$ and models with different complexities.
\subsection{Spatial independent, temporal independent}\label{sec:independent_time_location}
\begin{figure}[!htb]%
    \centering
    \subfigure[Plot of ${\bf z}_{tr}(t)$]{\includegraphics[width=0.48\textwidth, height=5cm]{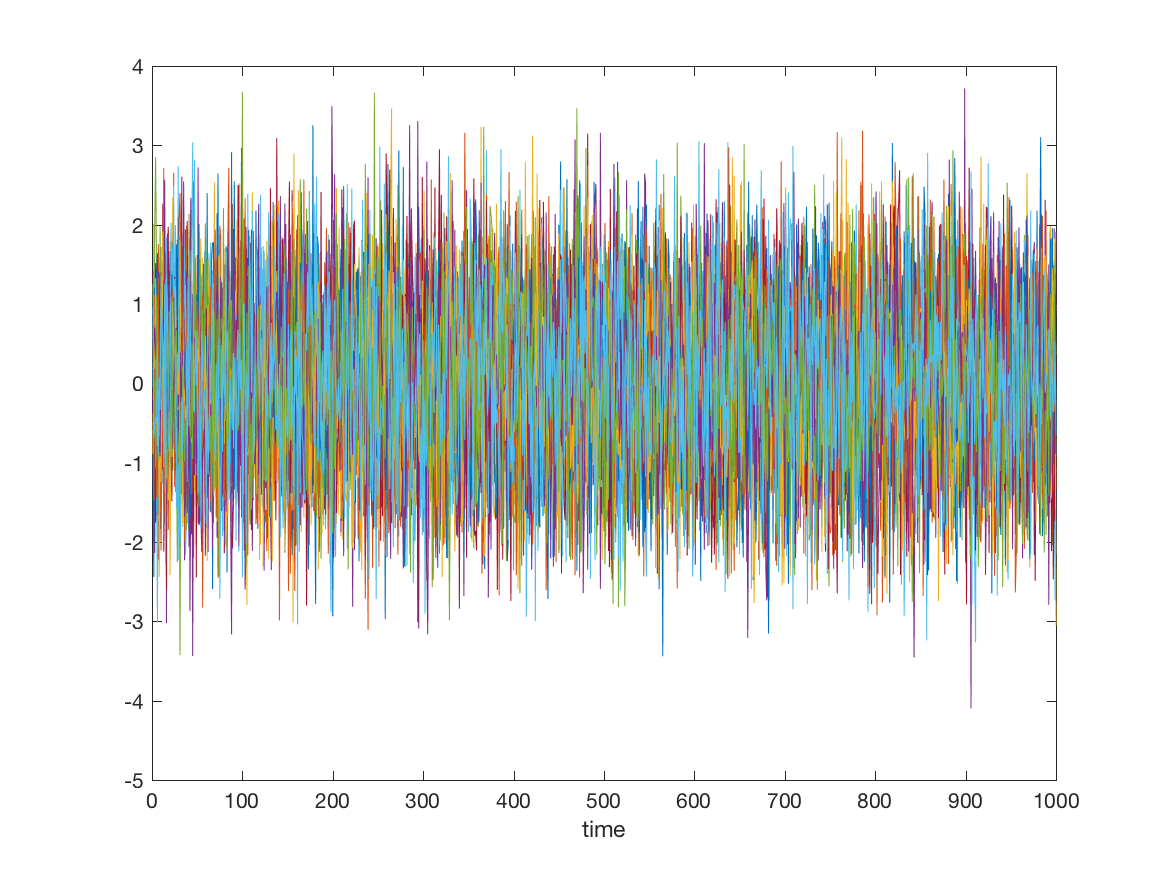} }%
    \subfigure[Reconstruction error]{\includegraphics[width=0.48\textwidth, height=5cm]{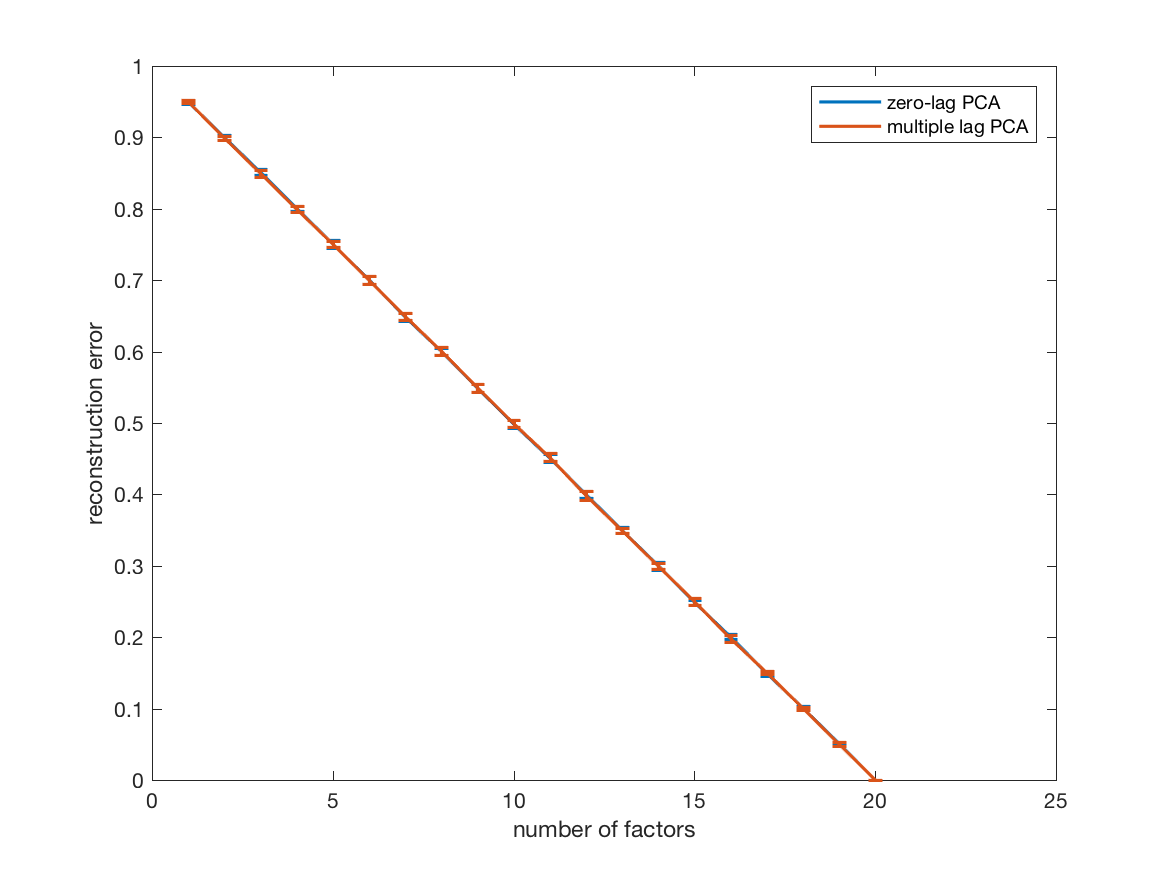}}%
    \caption{Independent time series. The time series ${\bf z}(t)$ has dimension $20$, for $t = 1, \dots, 1000$, ${\bf z}(t)$'s are iid random variables from multivariate Gaussian distribution with mean  zero and variance $I_{20}$. The reconstruction error (squared loss, i.e., the square of the Frobenius norm of the residuals) appears to be decreasing linearly when number of factors increases which means in the iid Gaussian case, all factors account for the same amount of the total variation.}%
    \label{fig:simu_1}%
\end{figure}

In this section, we consider the distribution of the data $F(Z)$ to be spatial independent and temporal independent. Specifically, the generated time series ${\bf z}(t)$ has dimension $20$, for $t = 1, \dots, 1000$, and that ${\bf z}(t)$'s are iid random variables from multivariate Gaussian distribution with mean  zero and variance $I_{20}$. It can be observed form Figure \ref{fig:simu_1} that comparing between two methods, the reconstruction errors evaluated on the test datasets are similar.  For both models, the reconstruct error appears to decrease linearly as the number of factors increases, which indicates that in the iid Gaussian case, all factors account for the same amount of the total variation. The result is as expected because (1.) for iid Gaussian random variable with identity covariance matrix, the projection on any direction will account for the same amount of variation and (2.) there is no lead-lag relationship between observations at different time points therefore in terms of predicting current observation, there is no gain of using observations from past or in the future. It can also be shown that in theory, two models will have the same solution when the signals are iid Gaussian with zero mean and identity covariance matrix.

\subsection{Temporal independent, spatially highly correlated}
\begin{figure}[!htb]%
    \centering
    \subfigure[Plot of ${\bf z}_{tr}(t)$]{{\includegraphics[width=0.33\textwidth]{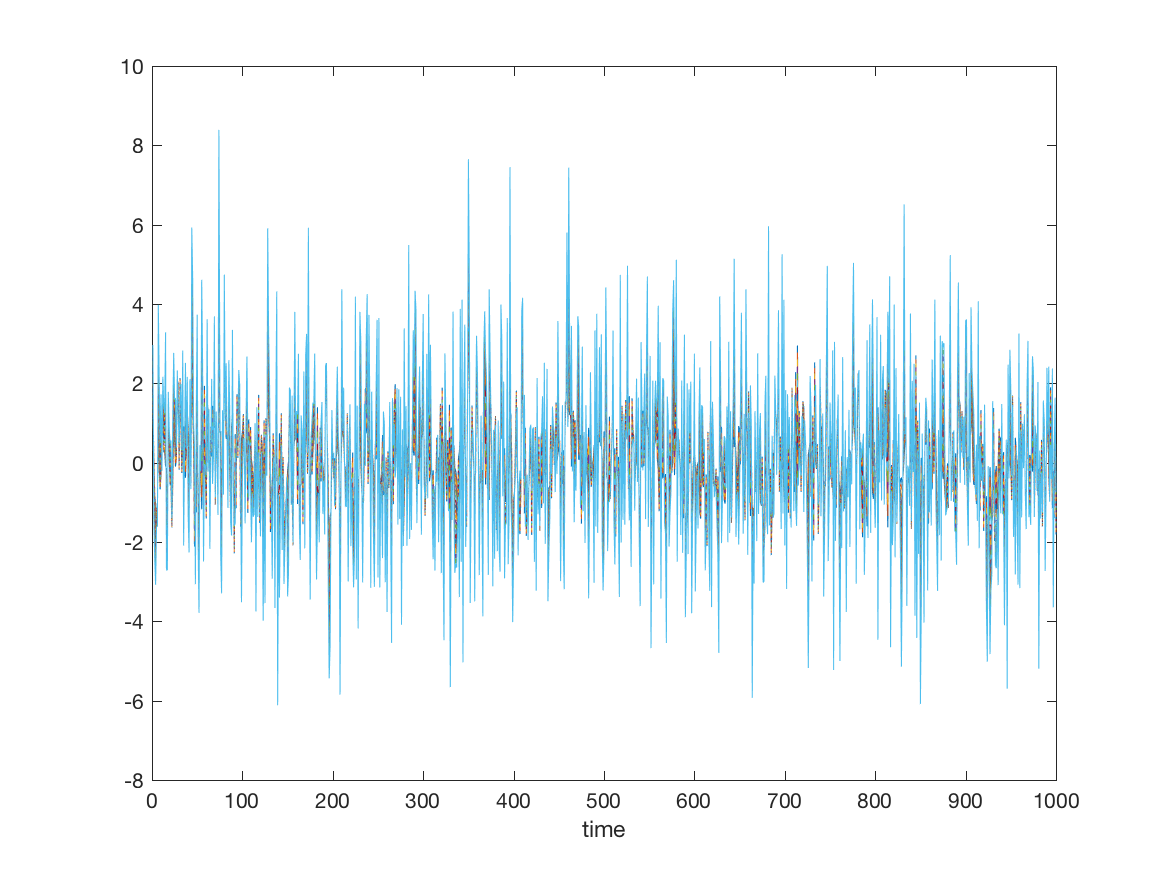} }}
      \subfigure[Correlation matrix of ${\bf z}(t)$]{{\includegraphics[width=0.33\textwidth]{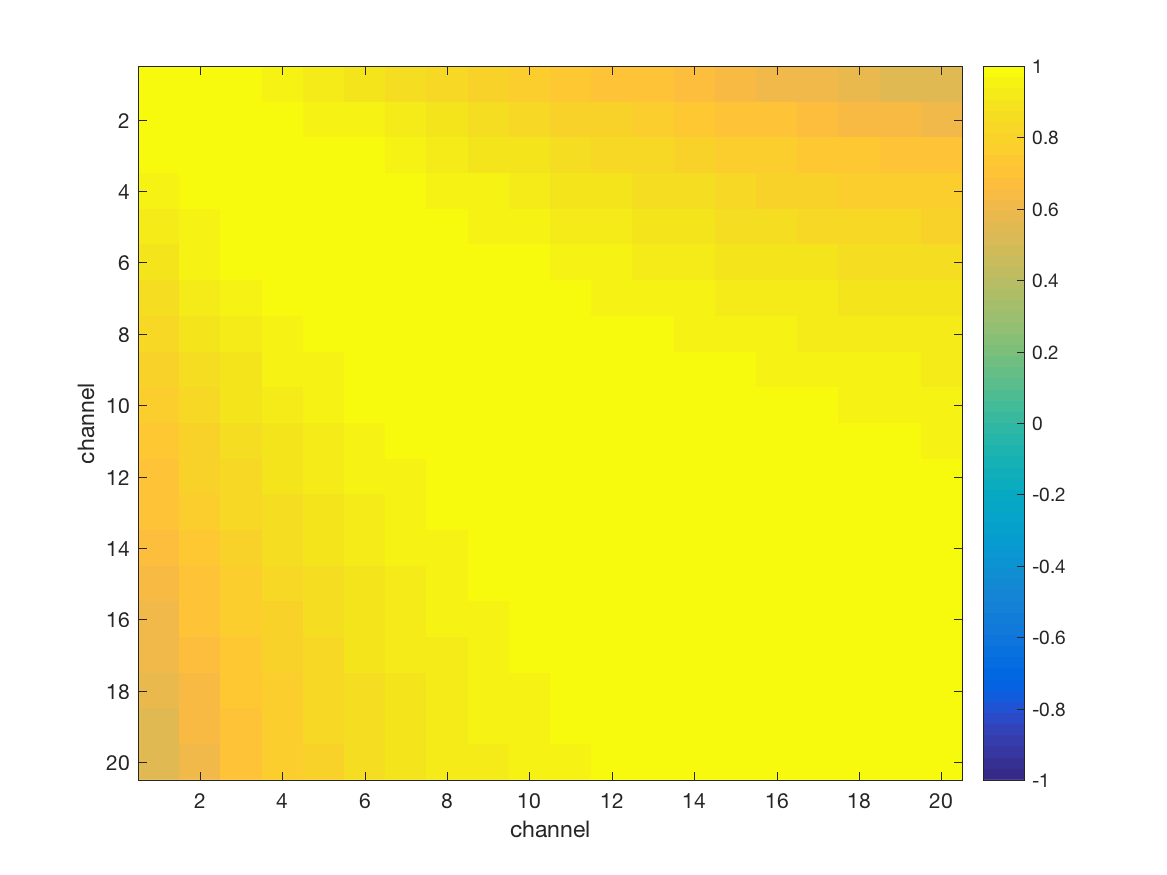} }}
    \subfigure[Reconstruction error]{{\includegraphics[width=0.33\textwidth]{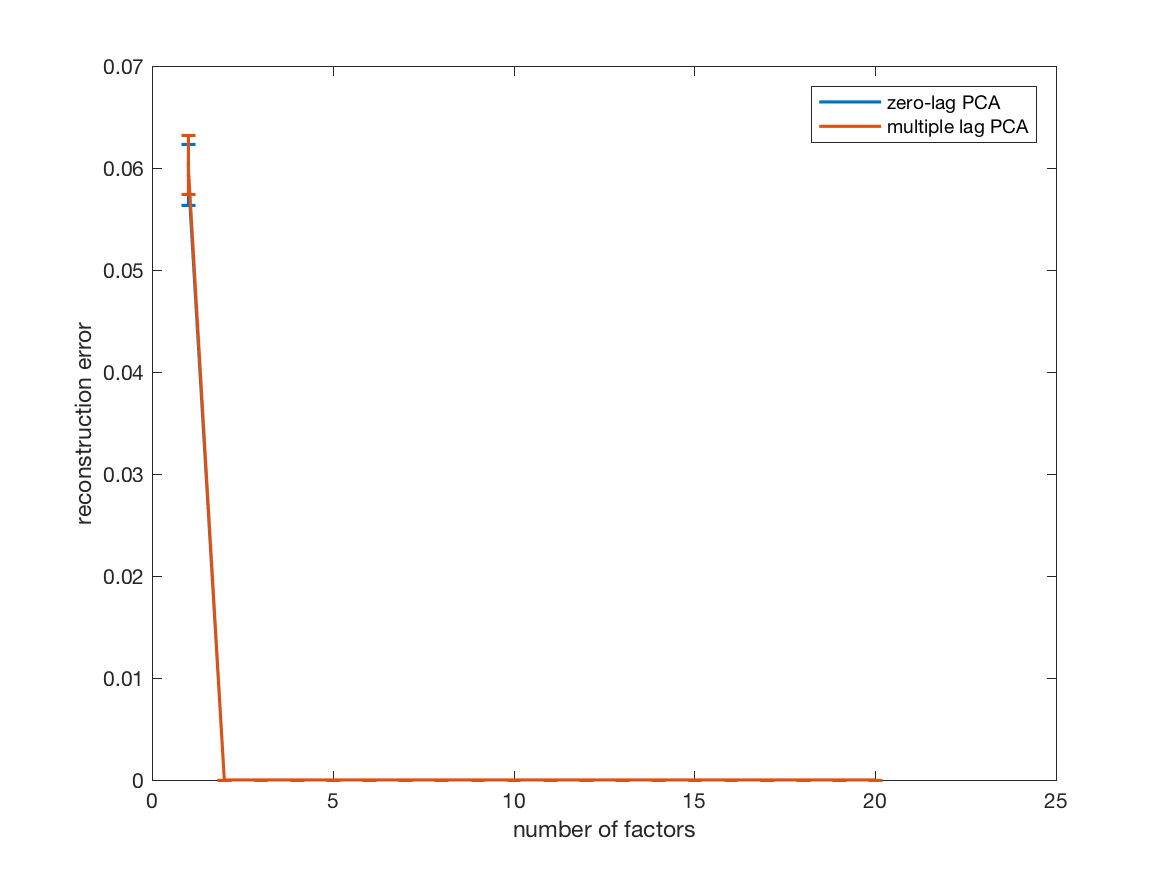} }}

    \caption{The time series data are generated such that they have highly spatial correlation but low temporal correlation (i.e, white noise). That is the correlation matrix $\Cor {\bf z}(t)$ has a low rank (rank is 2) structure and that ${\bf z}(t)$ is independent from ${\bf z}(t')$ for $t \neq t'$. The reconstruction error drops to zero when the number of factors is greater than 2, for both models. In this case, two models have similar reconstruction error evaluated on the test data.}%
    \label{fig:simu_2}%
\end{figure}

Similar to Section \ref{sec:independent_time_location}, we consider white noise time series where ${\bf z}(t)$ and ${\bf z}(t+h)$ are independent Gaussian variables for $h \neq 0$. In this simulation, the covariance $\Cov {\bf z}(t)$ has low rank (rank is 2), which means that at time $t$, the channels are highly correlated. Figure \ref{fig:simu_2} displays reconstruction error as a function of number of factors. It shows that two factors are capable of capturing all the dynamics of the input time series which is reasonable since since the generated time series has no temporal dynamics (they are temporally independent even though they have high spatial correlation). 

\subsection{Temporal correlated, spatially highly correlated}\label{sec:simulation_both_correlated}
\begin{figure}[!htb]%
    \centering
    \subfigure[Plot of ${\bf z}_{tr}(t)$]{{\includegraphics[width=0.48\textwidth, height=5cm]{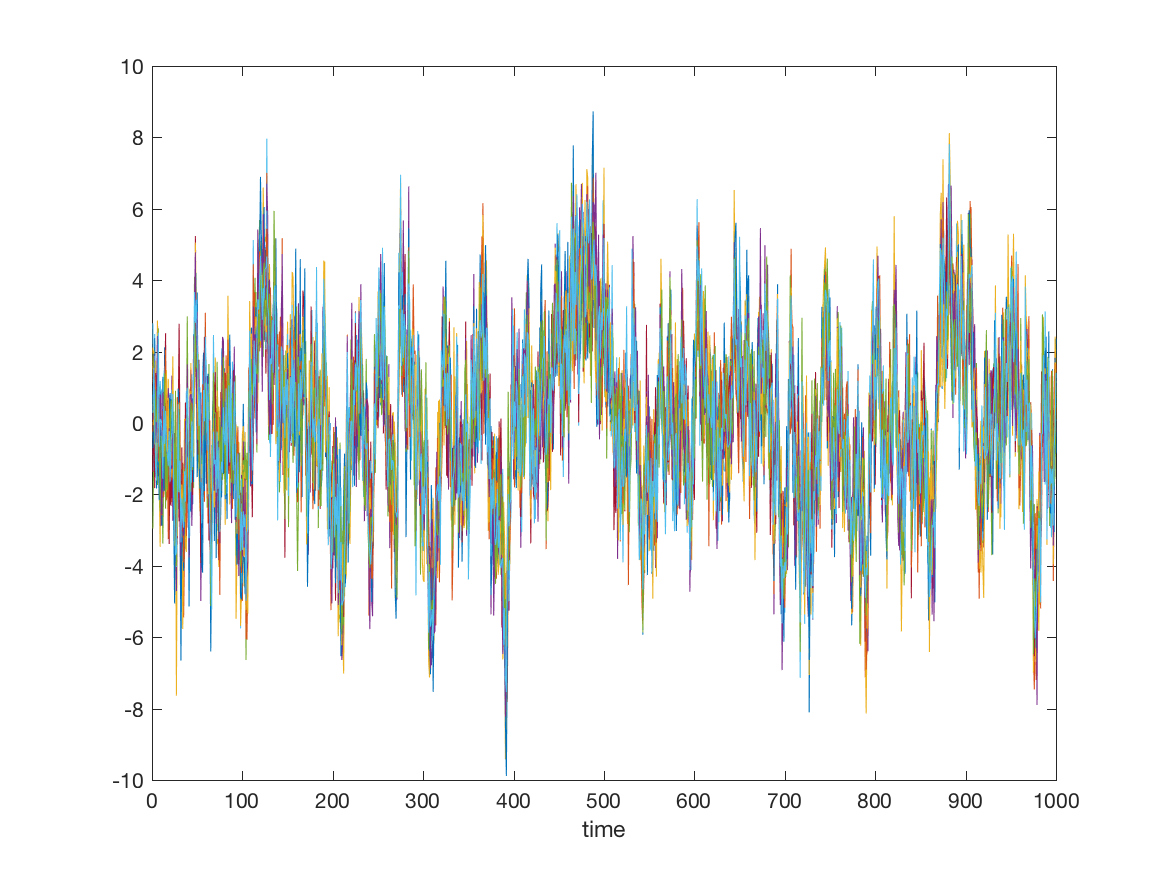} }}%
    \subfigure[Reconstruction error]{{\includegraphics[width=0.48\textwidth, height=5cm]{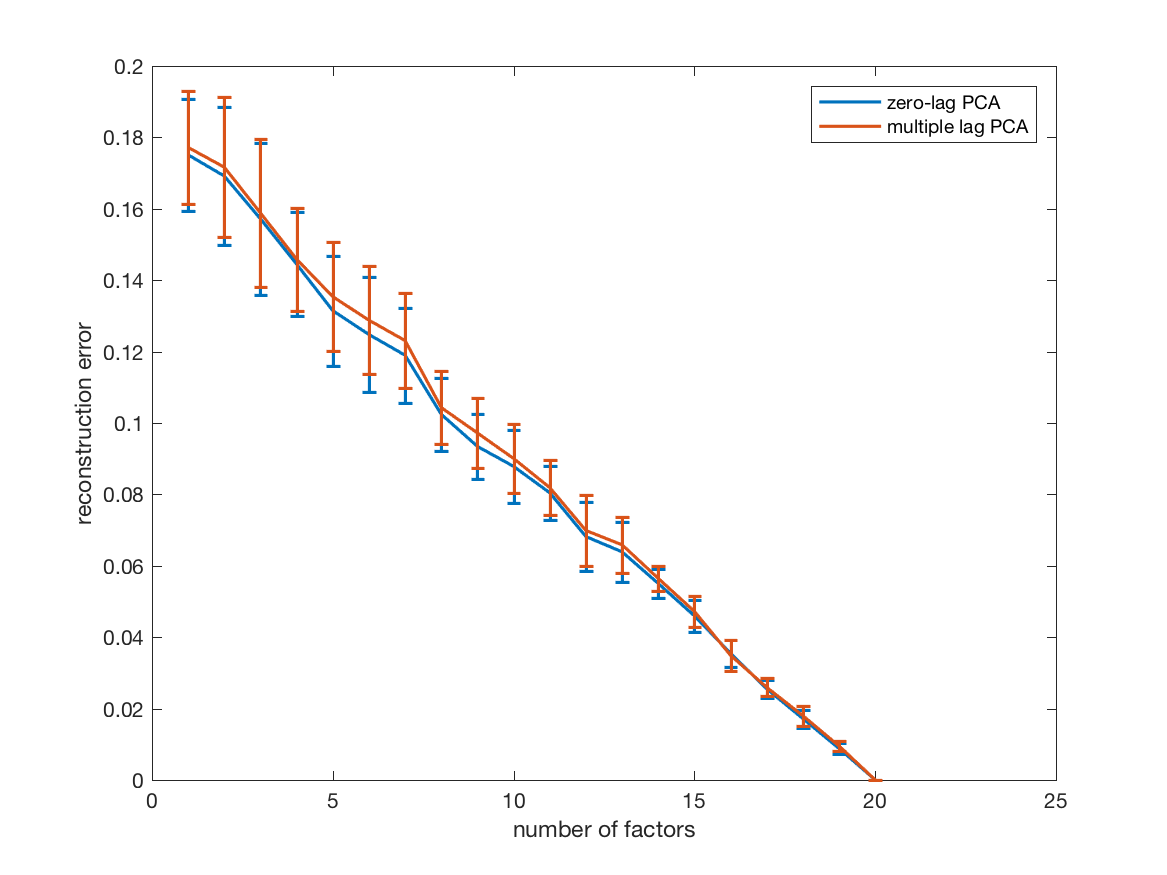} }}%
    \caption{The time series data are generated such that they have both strong temporal correlation and spatial correlation. The data is generated by first simulating data from autoregressive model $f_t = 0.9 f_{t-1} + \epsilon_t$ and then linearly projecting $f(t)$ to the observation space, which has dimension 20. Standard normal white noise has also been added to the observation. In terms of the reconstruction error on the test data, two models have similar performance. The variation of the reconstruction error shows a decreasing trend as number of factors increases. When the number of factors reaches 20, which is the dimension of the observation time series, the encoding $\rightarrow$ encoding procedure is equivalent to identity transformation.}%
    \label{fig:simu_3}%
\end{figure}

\begin{figure}[!htb]%
    \centering
     \subfigure[Plot of ${\bf z}_{tr}(t)$]{{\includegraphics[width=0.48\textwidth, height=5cm]{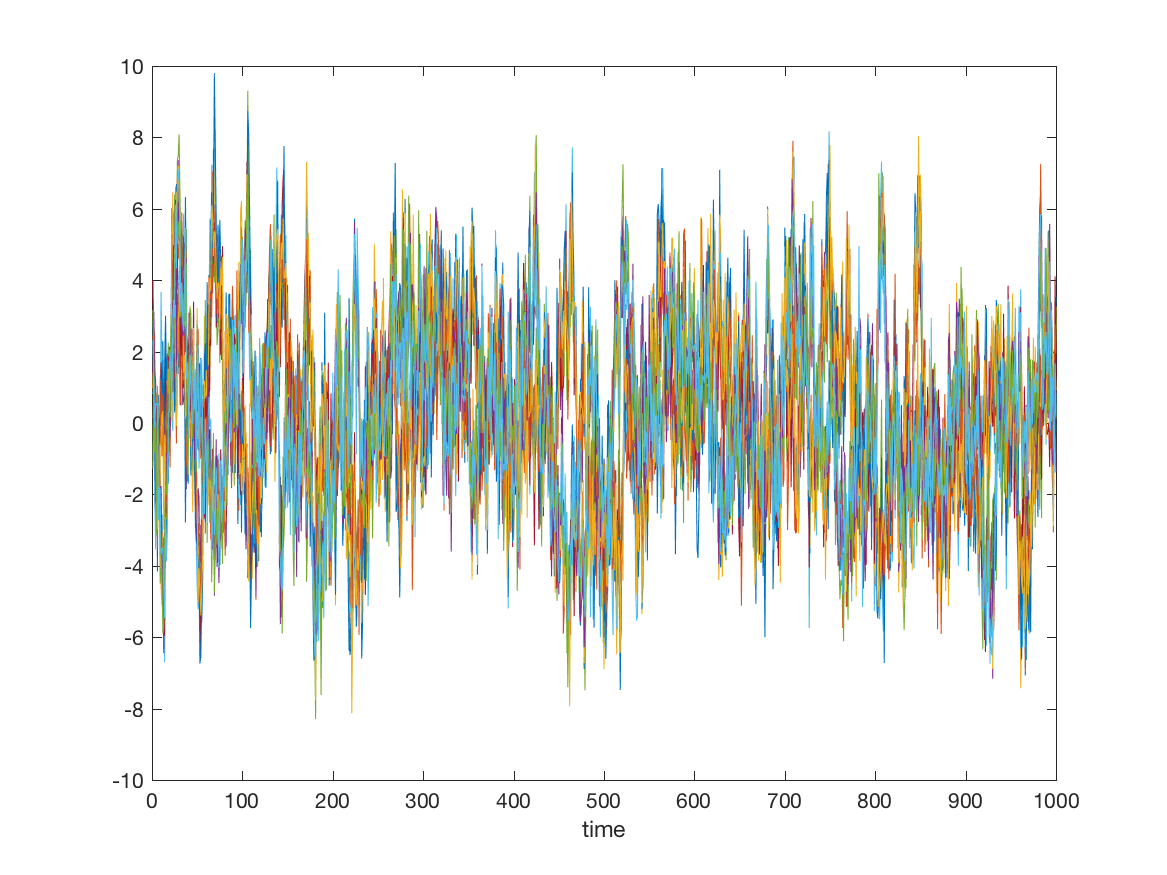} }}%
    \subfigure[Reconstruction error]{{\includegraphics[width=0.48\textwidth, height=5cm]{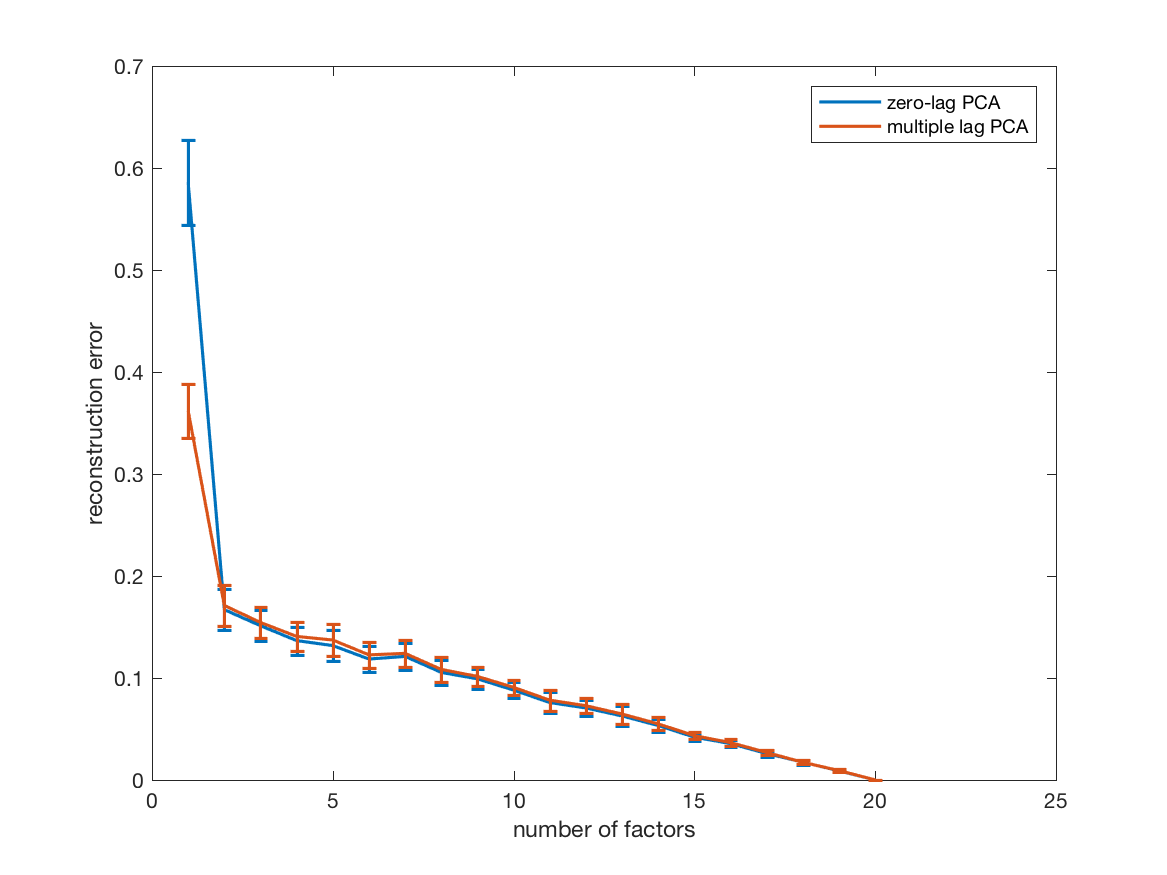} }}%
    \caption{Plot of shifted time series (two clusters) (left) and the reconstruction error evaluated on the test data (right). The time series is generated by first simulating time series ${\bf z}(t)$ following the same distribution as in Figure \ref{fig:simu_3}, and then shifting the first 10 channels by 40 time steps, i.e. $z_i(t) \leftarrow z_i(t+40)$ for $i = 1, \dots, 10$.  The plotted time series  (left) shows a two-cluster pattern. The reconstruction results show that the second model outperforms the first one when number of factors is 1 and that the two models have similar performance when the number of factors is greater than 1. The decreasing rate of the reconstruction error drops dramatically after the number of factors reaches the number of shifted time series clusters (2 in this case).}  %
    \label{fig:simu_4}%
\end{figure}

\begin{figure}[!htb]%
    \centering
     \subfigure[Plot of ${\bf z}_{tr}(t)$]{{\includegraphics[width=0.48\textwidth, height=5cm]{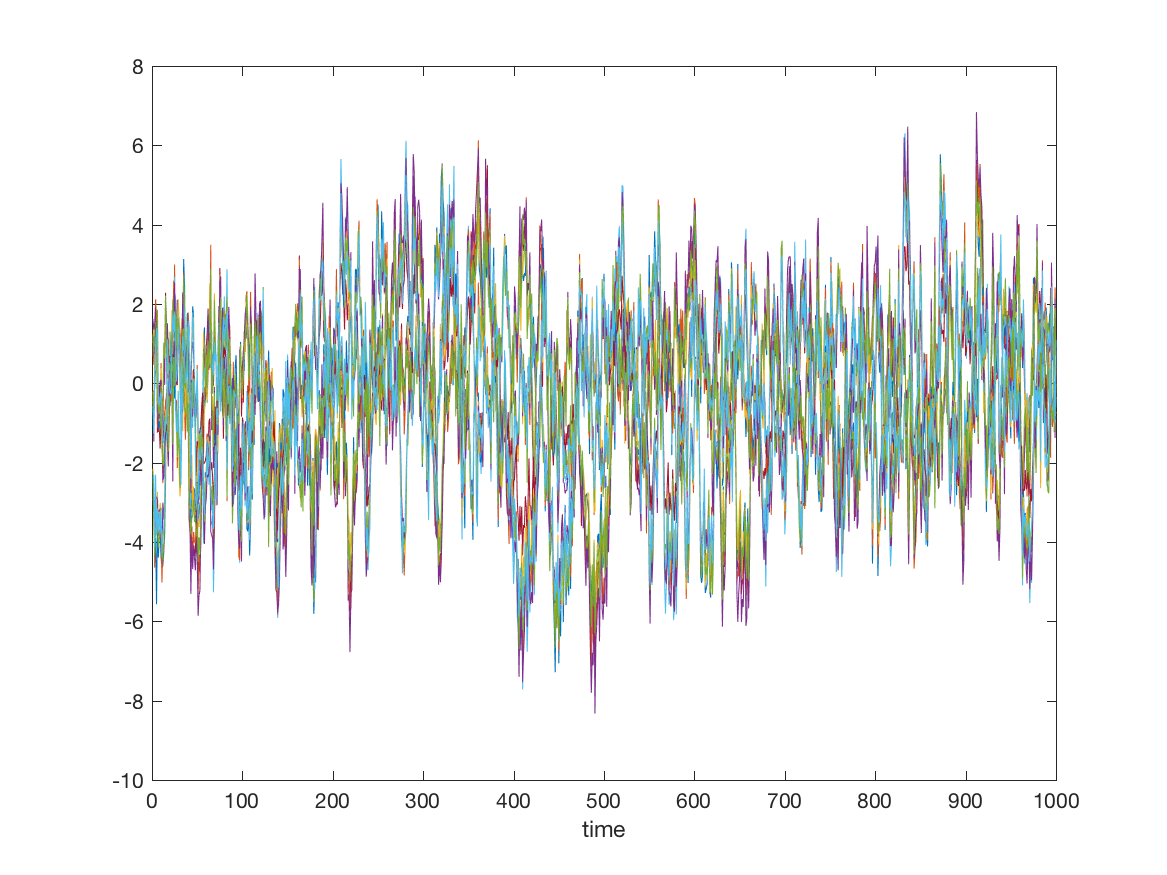} }}%
    \subfigure[Reconstruction error]{{\includegraphics[width=0.48\textwidth, height=5cm]{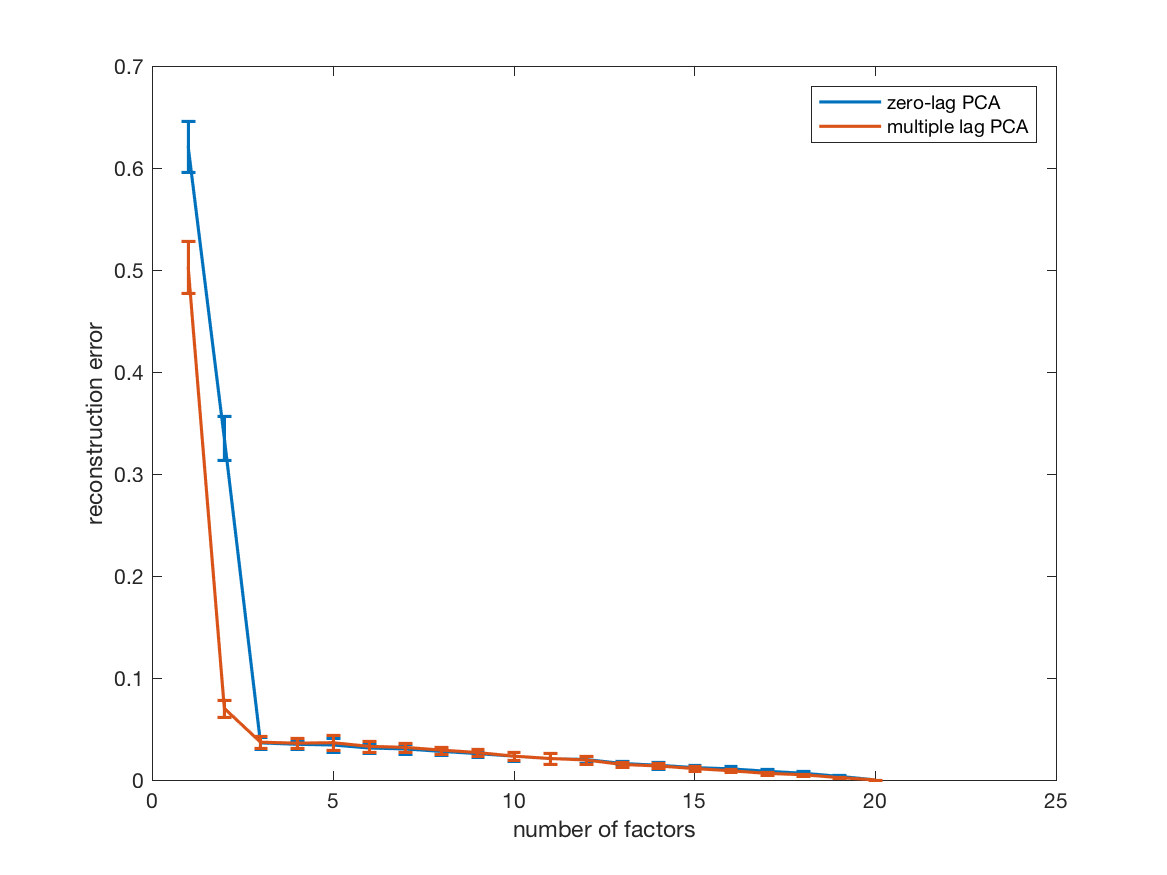} }}%
    \caption{Plot of shifted time series (three clusters) (left) and the reconstruction error evaluated on the test data (right). The time series is generated by first simulating  time series ${\bf z}(t)$ following the same distribution as in Figure \ref{fig:simu_3}, and then shifting the channels using $z_i(t) \leftarrow z_i(t+40)$ for $i = 1, \dots, 6$ and $z_j(t) \leftarrow z_j(t+80)$ for $j = 7, \dots, 12$. The plotted time series  (left) shows a three-cluster pattern. The reconstruction results show that the second model outperforms the first one when number of factors is smaller than 3 and that the two models have similar performance when the number of factors is greater than 3. The decreasing rate of the reconstruction error drops dramatically after the number of factors reaches the number of shifted time series clusters (3 in this case).}%
    \label{fig:simu_5}%
\end{figure}

In this simulation, we consider the time series ${\bf z}(t)$ that has both high spatial correlation and high temporal correlation. That is to say, ${\bf z}(t_1)$ and ${\bf z}(t_2)$ are correlated and ${z}_i(t)$ and ${z}_j(t)$ are also correlated. The time series is generated by first simulating data from autoregressive model $f_t = 0.9 f_{t-1} + \epsilon_t$ and then linearly projecting $f(t)$ to the observation space, which has dimension $20$. Standard normal white noises are then added to the observations. The training time series plot and the reconstruction errors are displayed in Figure \ref{fig:simu_3}. In terms of the reconstruction error on the test data, two models have similar performance. The variation of the reconstruction error shows a decreasing trend as the number of factors increases. When the number of factors reaches $20$, which is the dimension of the observations, the encoding-to-decoding procedure is equivalent to an identity transformation.

\subsection{Phase-shifted time series}
In this section we evaluate the model using the same data that is used in Section \ref{sec:simulation_both_correlated}, except that the time series from some of the channels are shifted. It is important to investigate shifts in time series because it is possible to have lead-lag relationships between EEGs in a region. The time series is generated by first simulating  time series ${\bf z}(t)$ following the same distribution as in Section \ref{sec:simulation_both_correlated}, and then shifting the channels. We perform two simulation studies, where in the first simulation, the time series is shifted  using $z_i(t) \leftarrow z_i(t+40)$ for $i = 1, \dots, 10$, and in the second simulation, the time series is shifted using  $z_i(t) \leftarrow z_i(t+40)$ for $i = 1, \dots, 6$ and $z_j(t) \leftarrow z_j(t+80)$ for $j = 7, \dots, 12$. 

It is observable from both Figure \ref{fig:simu_4} and \ref{fig:simu_5} that (1.) the time series plots show clearly clustered pattern, where within each cluster the signals are more synchronized, (2.) the second model, where the factor is a filtered version of the signal at all time points, gives a lower reconstruction error when the number of factors is smaller than the number of shifted clusters, and (3.) after the number of factors reaches the number of shifted time series clusters, the decreasing rate of the reconstruction error drops dramatically and the two models have similar performance.

The result is expected because model 2 is using information of all time lags to make prediction while model 1 only uses the instantaneous information, and hence in the shifted case, model 2 is more capable in capturing the temporal dynamics in the time series. The decreasing rate of the reconstruction error can be useful in estimating the number of synchronized clusters appeared in the data. 

\section{Exploratory analysis of the EEG data}\label{sec:data_analysis}
In this section, we perform exploratory analysis on real EEG data. The key challenge in analyzing EEG is the high dimensionality of the data. Computing dependence between regions or channels can be difficult due to the high dimensionality. Our goal here is address the dimensionality problem by deriving signal summaries (factors) of the EEGs in each region (e.g. SMA, left Pf) and then characterizing the dynamics and connectivity using the factor signals and the encoding/decoding functions.

The data were recorded during a motor learning study performed in the Stroke Rehab laboratory of our collaborator. The dataset contains EEG recordings for multiple subjects, where for each subject, 180 trails of 1 second EEG signals were recorded. The sampling rate of the data is 1000 Hz and number of channels is 256. The raw EEG data have been pre-processed by (1.) applying low pass filter at 50 Hz and (2.) using visual inspection and independent component analysis (ICA) to remove artifacts due to muscle activity, eye blinks and heart rhythms. Various analysis have been performed on the dataset, including using brain connectivity as predictor for ability of motor skill acquisitions  \citep{wu2014resting} and the analysis of curves of log periodograms using functional boxplots \citep{ngo2015exploratory}. In this paper, the goal of the exploratory analysis is to gain better understanding of the EEG data as well as the models that we used. 

\subsection{Computing regional summaries (factors)}
\begin{figure}[h]
    \centering
    \includegraphics[width=1.0\textwidth, height=10cm]{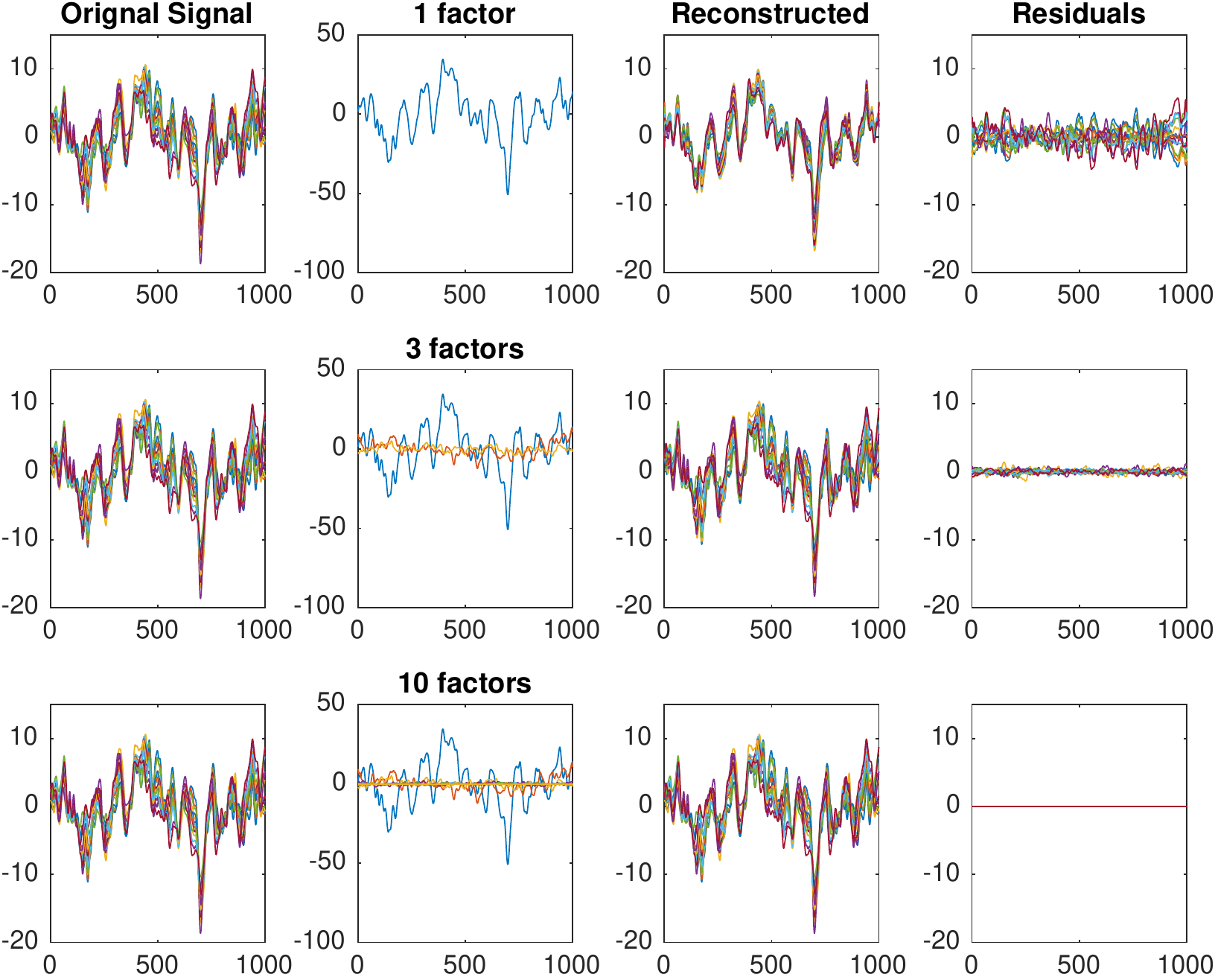}
    \caption{Signal compression using dynamic PCA (model 2). Column one shows the original EEG time series at SMA region, column two shows the factors computed via dynamic PCA, column three shows the reconstructed time series using different number of factors and column four shows the difference between the original signals and the reconstructed signals. Note that as the number of factors increases, the magnitude of the residuals decrease  (i.e., the squared error of reconstruction decreases).}
    \label{fig:Demo_signal_compression_specPCA}
\end{figure}

\begin{figure}[!htb]%
    \centering
     \subfigure{{\includegraphics[width=0.48\textwidth, height=5cm]{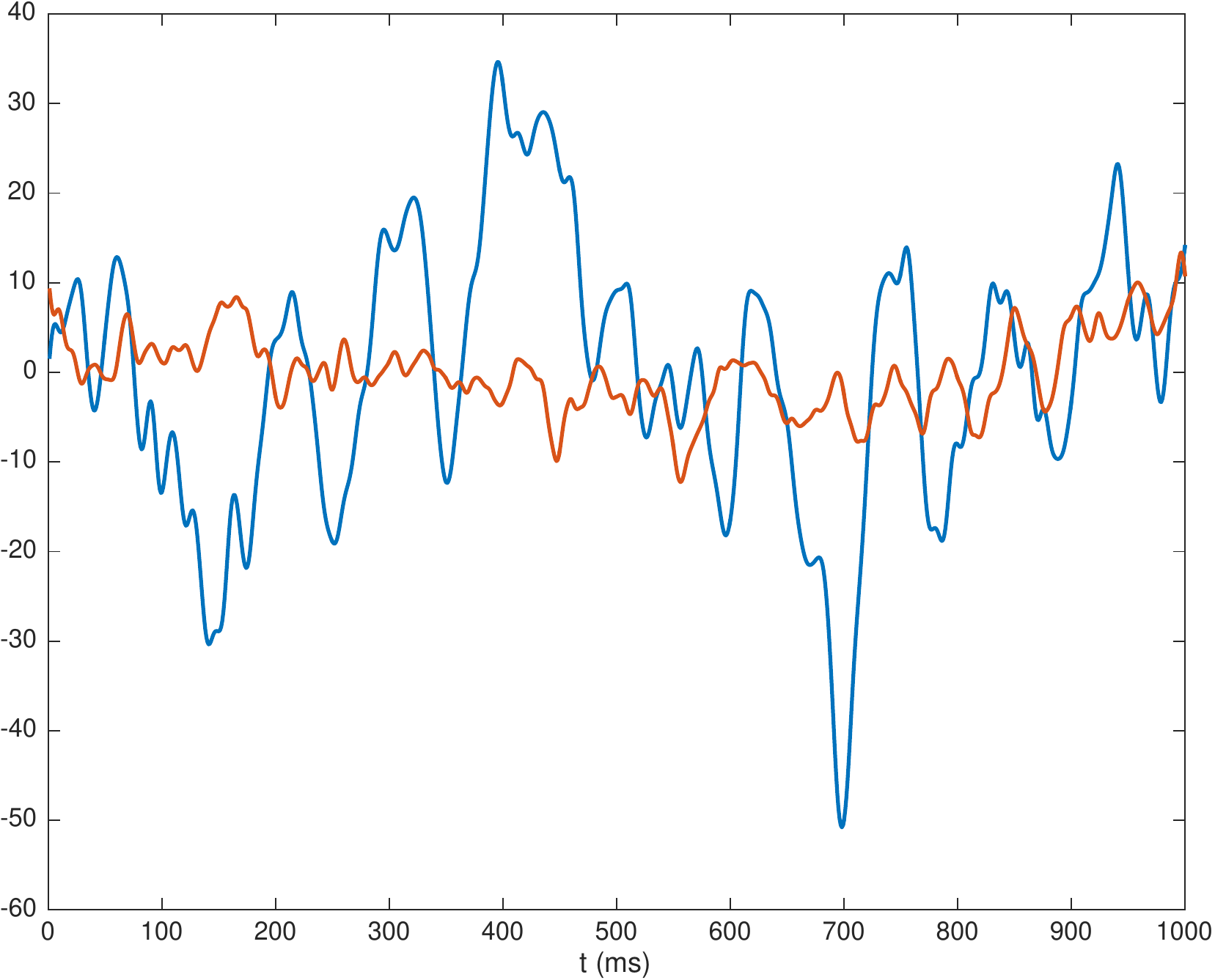} }}%
    \subfigure{{\includegraphics[width=0.48\textwidth, height=5cm]{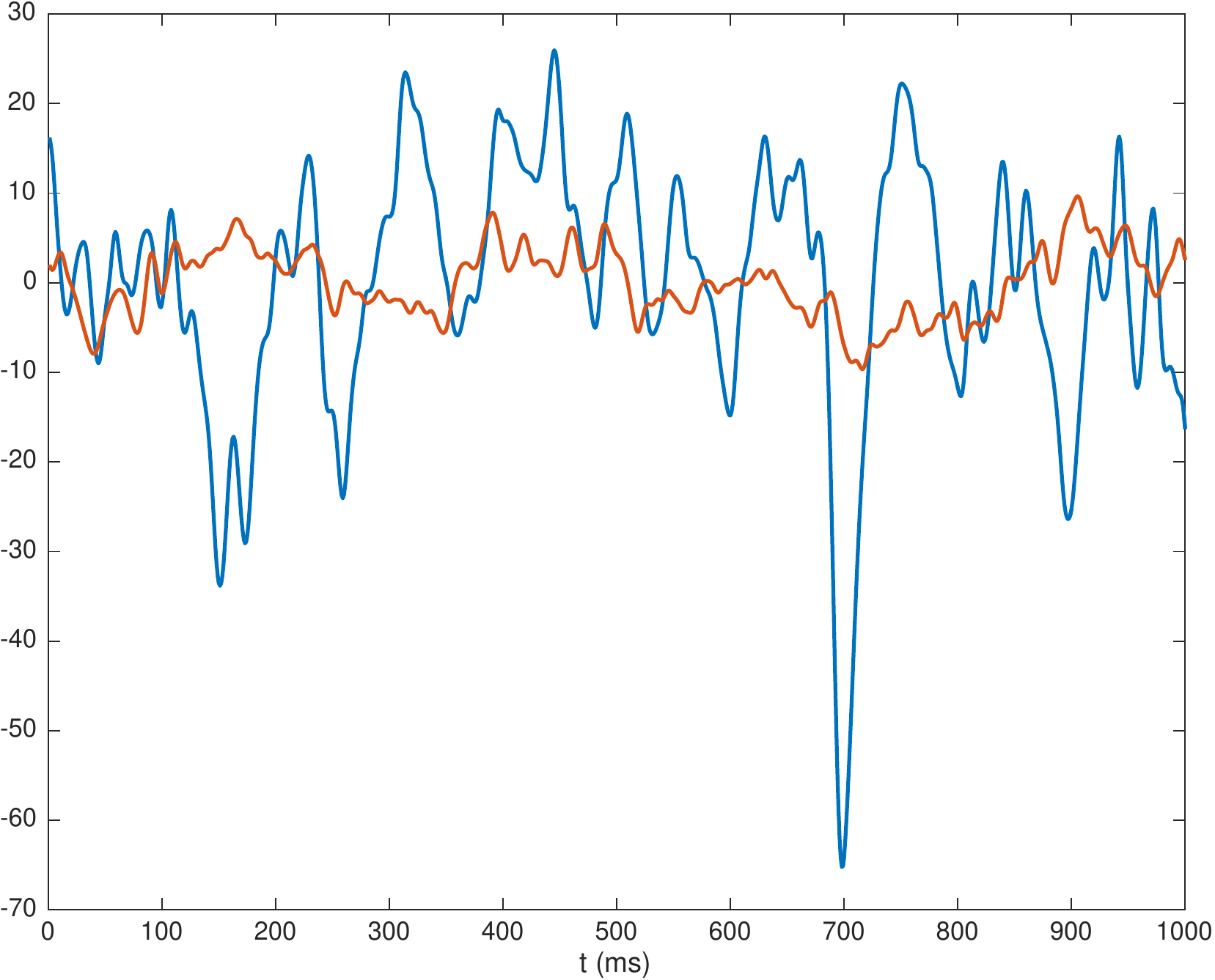} }}%
    \caption{Plot of the two factors in the SMA region (left) and the Left Pre-frontal cortex (right) that give the lowest squared reconstruction error.}%
    \label{fig:factor_time_series}%
\end{figure}

\begin{figure}[!htb]%
    \centering
     \subfigure{{\includegraphics[width=0.48\textwidth, height=5cm]{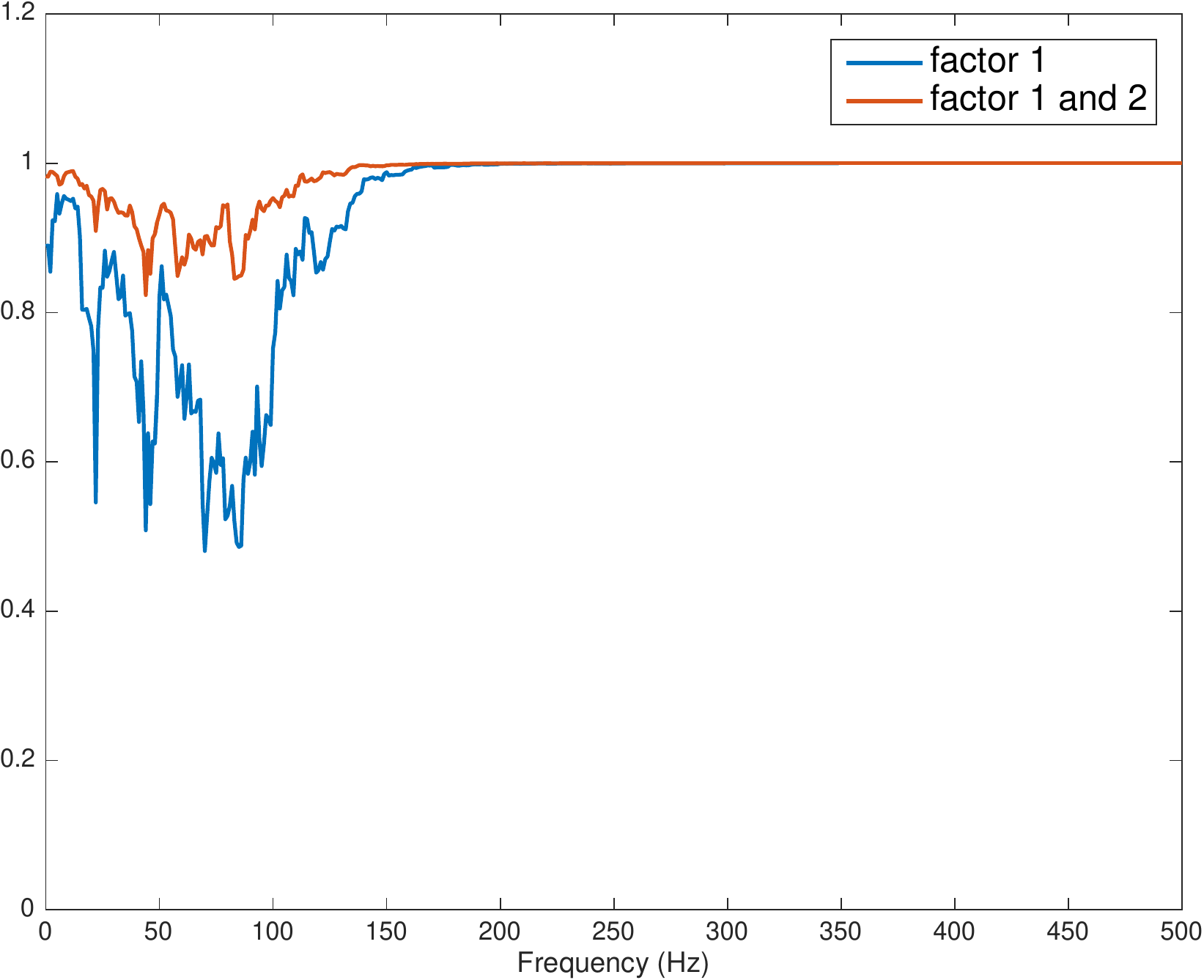} }}%
    \subfigure{{\includegraphics[width=0.48\textwidth, height=5cm]{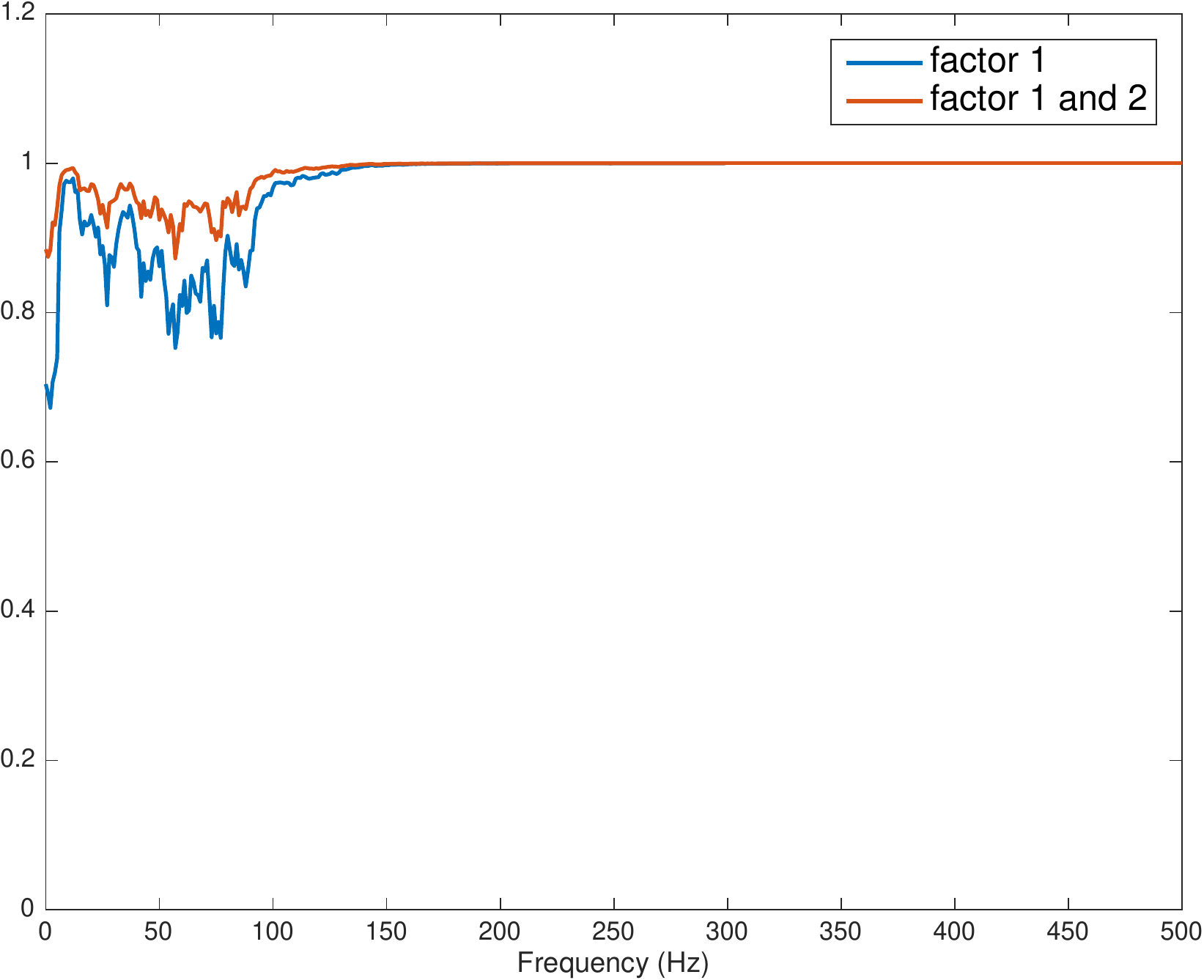} }}%
    \caption{Left: Variance accounted by factor 1 only and factors 1 and 2 in the SMA region. Right: Variance accounted by factor 1 only and factors 1 and 2 in the Left Pre-frontal region. Note that two factors explain around 90\% of the total variance in the high dimensional EEGs. Or equivalently these 2 factors constraint the reconstruction error to be under 10\%. }
    \label{fig:var_accounted_by_factors}%
\end{figure}

Figure \ref{fig:eeg_two_regions} displays the EEG signals recorded for one subject at one trial. It can be observed that the EEG signals are very highly synchronized, which means EEG at channel $i$ is highly correlated with EEG at channel $j$ at the same time $t$. It also appears that signals within the same region (e.g. SMA and left Pr) have higher correlation, comparing to that of the signals in different regions. Due to these high spatial correlations, it is sensible to represent these EEGs in terms of low dimensional summaries that capture the most variation in these EEG signals. Figure \ref{fig:Demo_signal_compression_specPCA} shows the reconstruction of EEGs at SMA region using the linear convolution encoder. It can been seen that as the number of factors increases, the magnitude of the residuals decrease. The top two summary signals (factors) computed using EEGs from SMA region and left Pre-frontal region are plotted in Figure \ref{fig:factor_time_series} and the proportion of total variation accounted by theses factors are shown in  Figure \ref{fig:var_accounted_by_factors}. The results for both regions show that factors with very low dimension (less than 3) can represent most of the variation of the original signals. This is consistent with the fact that the EEGs are highly correlated spatially due to volume conduction.

\subsection{Properties of the summary factors}
\begin{figure}[!htb]%
    \centering
     \subfigure{{\includegraphics[width=0.48\textwidth, height=5cm]{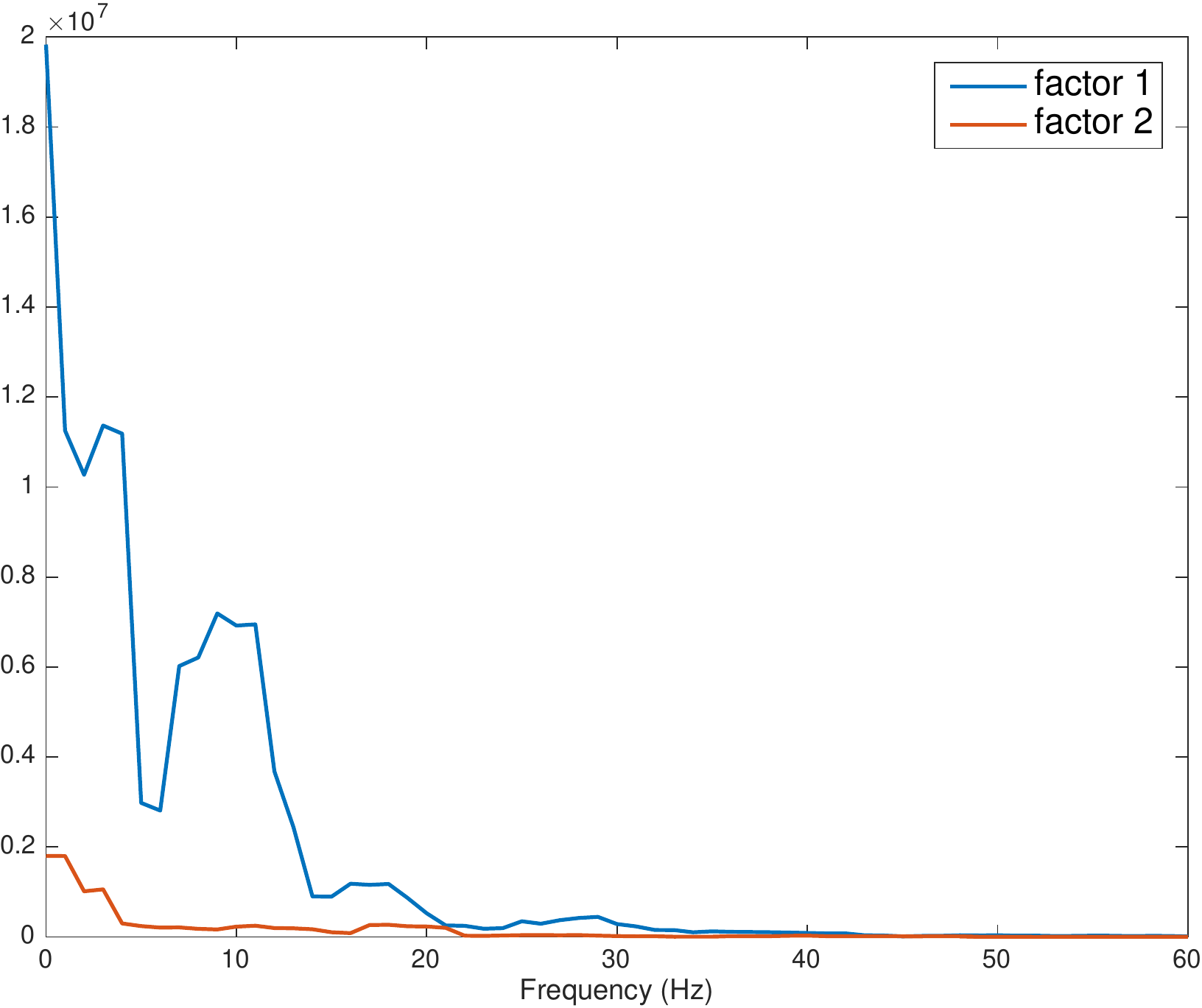} }}%
    \subfigure{{\includegraphics[width=0.48\textwidth, height=5cm]{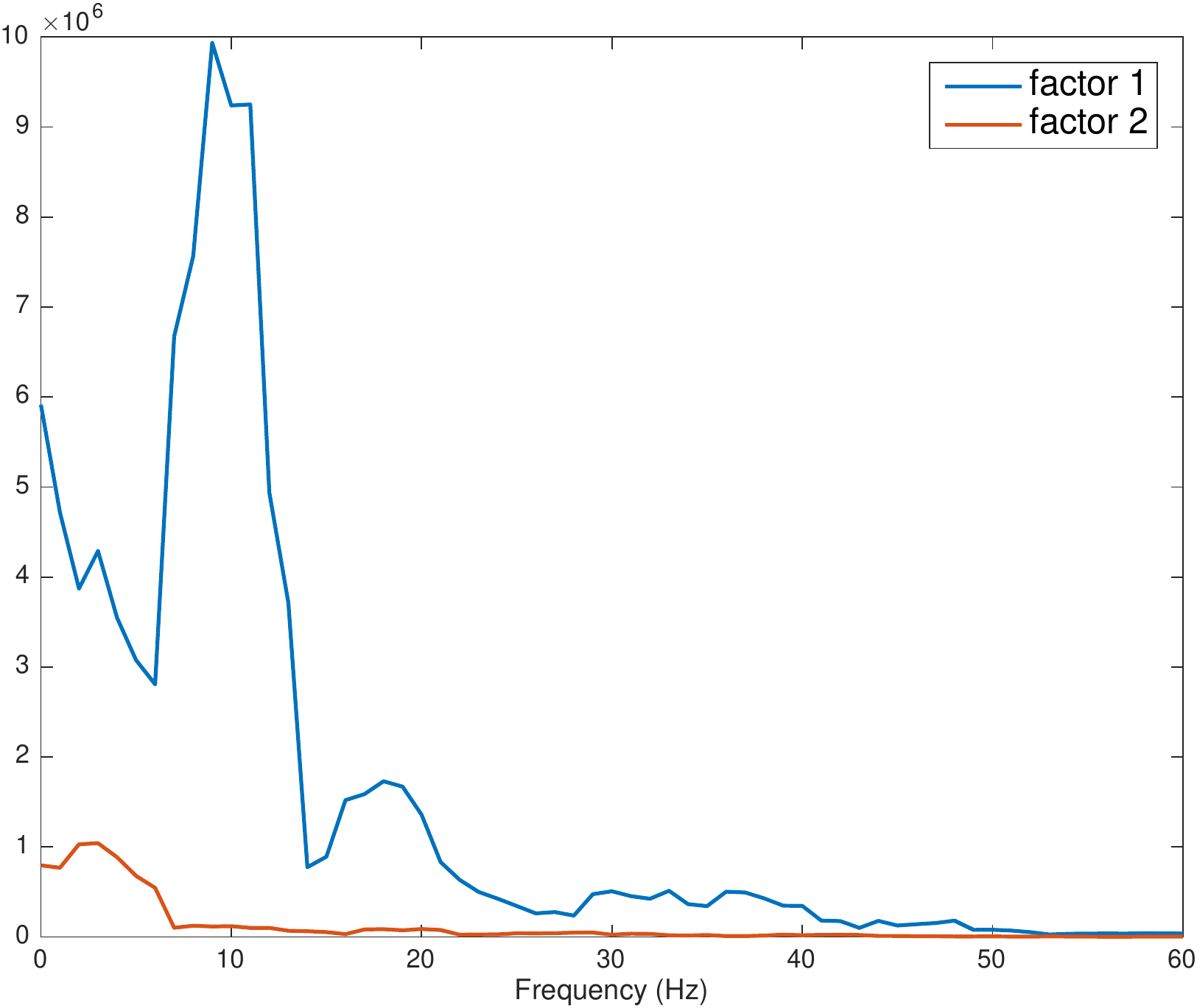} }}%
    \caption{Estimated power spectrum for the first two factors in the SMA region (left) and the Left Pre-frontal region (right).  Factor 1 in both the SMA and Left Pre-frontal regions capture the alpha oscillations (8-12 Hertz) and low beta (16-30 Hertz). Factor 2 accounts for the delta and theta band oscillations (1-8 Hertz). }%
    \label{fig:power_spectrum_density_of_factors}%
\end{figure}

\begin{figure}[h]
    \centering
    \includegraphics[width=1.0\textwidth, height=10cm]{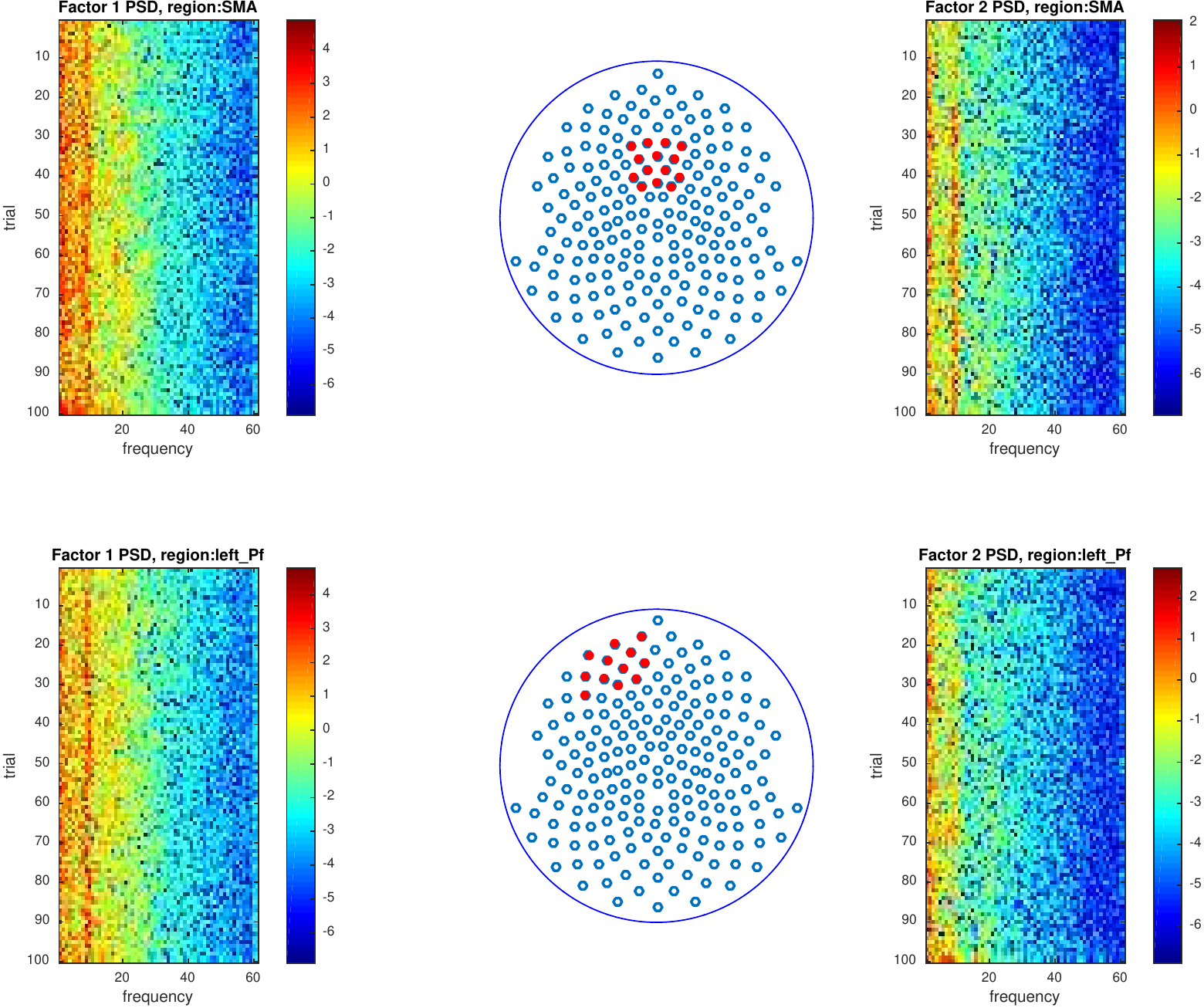}
    \caption{Top: SMA region. Estimated power spectrum across 100 epochs of the first factor (left) and the second factor (right). Bottom: Left Pre-frontal region. Estimated power spectrum across 100 epochs of the first factor (left) and the second factor (right).}
    \label{fig:power_spectrum_density_of_factors_multiple_trials}
\end{figure}

\begin{figure}[h]
    \centering
    \includegraphics[width=1.0\textwidth, height=10cm]{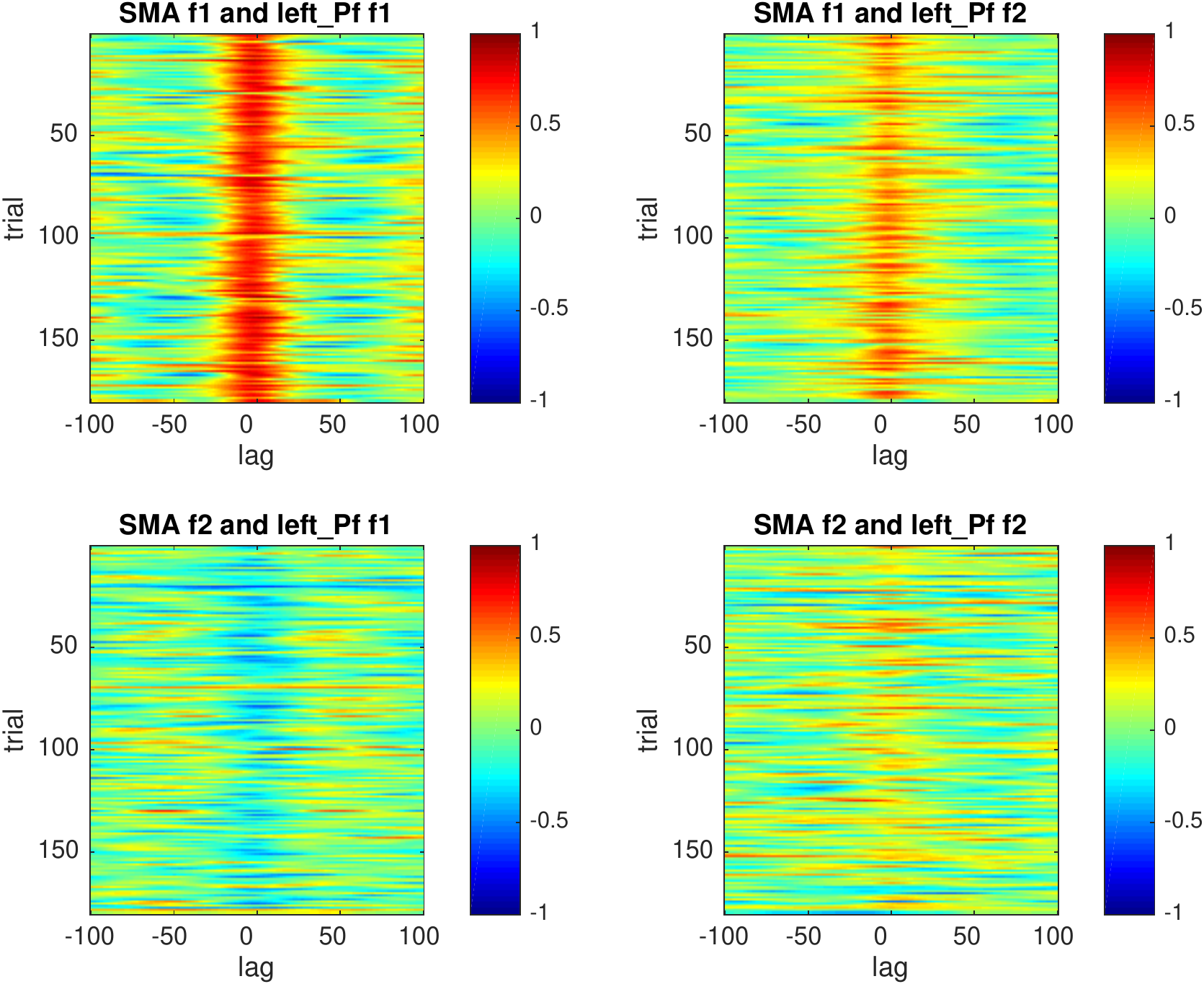}
    \caption{Cross correlation between factors in SMA region and Left Pre-frontal region across 180 epochs. Top left: correlation between factor 1 in SMA region and factor 1 in Left Pre-frontal region; top right: correlation between  factor 1 in SMA region and factor 2 in Left Pre-frontal region; bottom left: correlation between factor 2 in SMA region and factor 1 in left pre-frontal region; bottom right: correlation between factor 2 in SMA region and factor 2 in left pre-frontal region. The correlations show consistent patterns across epochs. }
    \label{fig:factors_cross_correlation_multiple_trial}
\end{figure}

Figure \ref{fig:power_spectrum_density_of_factors} shows the estimated power spectrum density of the top factors computed for SMA region and left Pf region. It shows that factor 1 in both the SMA and Left Pre-frontal regions capture the alpha oscillations (8-12 Hertz) and low beta (16-30 Hertz). Factor 2 has more power in the delta and theta band oscillations (1-8 Hertz). The power spectrum across 100 EEG epochs are estimated and visualized in Figure \ref{fig:power_spectrum_density_of_factors_multiple_trials}. The results  show that the spectrum pattern for the top factors are consistent across trails, where factor 1 concentrates more on alpha oscillations (8-12 Hertz) and  factor 2 concentrates more on the delta and theta band oscillations (1-8 Hertz). In order to study the temporal dependence between the factors, we plot the cross-correlation between the top two factors, evaluated for multiple epochs (Figure \ref{fig:factors_cross_correlation_multiple_trial}). The cross-correlation between factor 1 in SMA region and factor 1 in left Pre-frontal region appears to be very consistent across epochs. The cross-correlation that involves factor 2 also shows some consistent patterns across epochs, although the consistency is weaker comparing to the cross-correlation between factor 1's in two regions.

\subsection{Interactive Matlab toolbox for exploratory analysis}
\begin{figure}[!htb]
    \centering
    \includegraphics[width=1.0\textwidth, height=10cm]{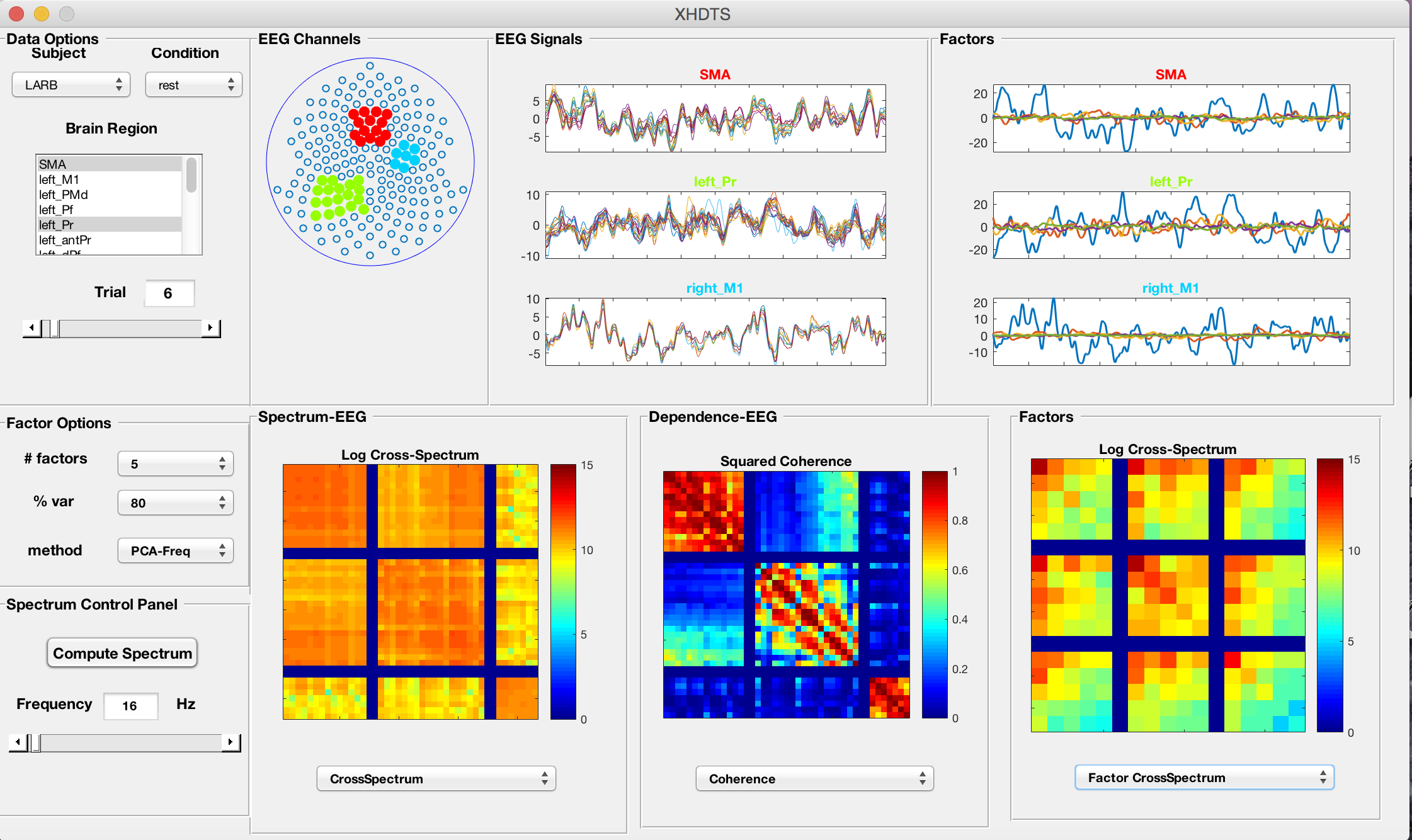}
    \caption{The interface of the XHiDiTS Toolbox: exploratory high dimensional time series toolbox in Matlab. This is an interactive toolbox where the user selects the dataset 
to be analyzed. From the dataset, the user can select specific regions of interest (ROIs) to be 
analyzed. This toolbox supports a rich set of options and methods for visualizing and analyzing high dimensional time series, including the methods presented in this paper. This toolbox is actively developed and maintained. It can be downloaded from {https://goo.gl/uXc8ei}.}
    \label{fig:matlab_toolbox}
\end{figure}

We implemented and actively maintain a Matlab toolbox (Exploratory High-Dimensional Time Series (XHiDiTS) toolbox https://goo.gl/uXc8ei) with a graphical interface that allows users to performance exploratory analysis easily. Figure \ref{fig:matlab_toolbox} shows a screen shot of the toolbox interface.

The option panels provide a rich set of options that allow users to select from by just one-clicking. The options include  (1.) subject-specific data; (2.) experimental conditions (resting state vs task); (3.) specific regions (users can load their own channel location/ grouping files) (4.) methods for learning lower dimensional representations and the complexity of the model (e.g. number of factors); and (5.) methods for computing connectivity (e.g. partial directed coherence, correlation matrix, coherence matrix and block coherence). 

The visualization panels show (1.) the 2-d scalp, with selected regions highlighted and colored, where the coloring is consistent with the title of the signal plot, allowing users to match the plot and region easily; (2.) the signals and factors (low dimensional representations) for selected regions; and (3.) the spectrum of the signals and the connectivity maps.  

The toolbox has low latency in updating the results for datasets with reasonable sizes. For example, for a 256-channel EEG data that contains 1000 time points for each channel and 200 epochs, the latency for updating the results for a new setup is within seconds (<1s for most of the methods). Users can also load their own datasets or add their own definition of functions for connectivity and other quantities. 

\section{Conclusion and future work}\label{sec:conclusion}
In this paper, we developed exploratory procedures for high dimensional EEGs under the presence of high multi-collinearity by using low-dimensional representations.  We evaluated (benchmark) the performance of the dimension-reduction methods via numerical experiments by applying the models on time series generated form different distributions, thus provided guidelines for the application scope of the methods. We performed exploratory analysis on a real EEG dataset to gain deeper understanding of the methods. The results for both of the simulation and exploratory analysis show that learning low-dimensional representations (factors) has potential benefits for subsequent modeling of the connectivity in high dimensional time series because the factors are capable of preserving the dynamics of the data (i.e., temporal dynamics,  variation) while reducing dimension (complexity) of the original problem. We also implemented the methods in a Matlab toolbox with graphical interface that allows users to interactively explore, process and analyze the data in a convenient way.

Our future work in this area includes a comprehensive evaluation of the methods. For example, we would like evaluate the ability of the models in capturing the temporal dynamics and at the same time, quantify the artifact that might be induced by the mixing. Moreover, to make the package more comprehensive, we shall include other emerging measures of dependence such as isolated coherence \citep{pascual2014isolated, ombao2008evolutionary, fiecas2011generalized, yuxiao2016bookchapter} and other more general (possibly non-linear) methods for obtaining summary signals \citep{pe2015generalized}.

%%%%%%%%%%% BIBLIOGRAPHY %%%%%%%%%%%%%%%%%%
\bibliography{ywang}{}
\bibliographystyle{apalike}

\clearpage
\newpage
\begin{appendices}

\section{Algorithms}
\begin{algorithm}
\caption{Compute Spectral Matrix}
\label{algo:spectrum}
\begin{algorithmic}[1]
\fontsize{9.4pt}{9.4pt}\selectfont
\Procedure{CrossPowerSpectrum}{\{${\bf z}(t)$\}}
\State set nfft = $T$, $m = [\sqrt T]$, $h_\ell = \frac{1}{2m+1}$ 
\State //Remark 1: $h_{\ell} \geq 0$, $h_{-\ell} = h_{\ell}$ and $\sum_\ell h_\ell = 1$
\State //Remark 2: In order to make $\widehat{S}_{zz}$ positive definite we need $2m+ 1 > p$
\For {$k = 0, \dots, T-1$}
%\State ${\bf x}(k) = \sum_{t=1}^{T}$
\State compute  ${\bf z}_{\omega}(k) \gets \sum_{t=0}^{T-1}{\bf z}(t)\exp(-i2\pi t \frac{k}{T})$ \Comment transfer function: fft
\State compute $I_{\omega}(k) \gets {\bf z}_{\omega}(k){\bf z}^*_{\omega}(k)$ \Comment raw periodogram
\EndFor
\For {$k = 0, \dots, m$} \Comment padding 
\State $I_{\omega}(-k) \gets I^*_{\omega}(k)$
\State $I_{\omega}(T+k) \gets I^*_{\omega}(T-k)$
\EndFor

\For {$k = 0, \dots, T-1$} \Comment smoothing
\State compute $\widehat{S}_{zz}(k) \gets \sum_{\ell=-m}^{m} h_\ell I_{\omega}(k+\ell)$ 
\EndFor

\Return  \{${\bf z}_\omega(k)$\}, \{$\widehat{S}_{zz}(k)$\}
\EndProcedure

\end{algorithmic}
\end{algorithm}

\begin{algorithm}
\fontsize{9.4pt}{9.4pt}\selectfont
\caption{Factor model - $m$ components}
\label{algo:pca_freq}
\begin{algorithmic}[2]
\Procedure{PCA}{\{$\widehat{S}_{zz}(k)$\}}
\For {$k = 0, \dots, T-1$} 
\State compute eigen values $\lambda_1(k) > \lambda_2(k), \dots, \lambda_p(k)$
\State compute corresponding eigen vectors ${\bf e}_1(k), {\bf e}_2(k), \dots, {\bf e}_p(k)$
%\State sd
\EndFor

\Return \{$\lambda_j(k)$\}, \{${\bf e}_j(k)$\}
\EndProcedure

\Procedure{ComputeFactors}{\{${\bf z}_\omega(k)$\}, \{$\lambda_j(k)$\}, \{${\bf e}_j(k)$\}}
\For {$k = 0, \dots, \lfloor T/2 \rfloor $}
\State ${\bf C}(k) \gets [{\bf e}_1(k), \dots, {\bf e}_m(k)]$
\EndFor

\For {$k = \lfloor T/2 \rfloor + 1, \dots, T-1$}
\State ${\bf C}(k) \gets {\bf C}^*(T-k)$
\EndFor

\For{$k = 0, \dots, T-1$}
\State ${\bf f}_\omega(k) = {\bf C}^* (T-k){\bf z}_\omega(k)$ \Comment transfer function for ${\bf f}(t)$ 
\EndFor

\For {$t = 0, \dots, T-1$}
\State ${\bf f}(t) = \frac{1}{T} \sum \limits_{k=0}^{T-1} {\bf f}_\omega(k) \exp(i2\pi t \frac{k}{T})$ \Comment ifft 
\EndFor

\Return \{${\bf f}(t)$\}, \{${\bf C}(k)$\}
\EndProcedure

\end{algorithmic}
\end{algorithm}

% Figures
\clearpage
\newpage
\section{Figures}

% The figures

\end{appendices}

\end{document}